\def\eiso{E_{\rm iso}}

\def\ep{E_{\rm peak}}
\def\eop{E^{\rm obs}_{\rm peak}}
\def\se{S_E}

\def\Fp{F_{\rm peak}}

\def\fn{F^{\rm P}_N}

\def\bsax{{\it Beppo}SAX}

\documentclass[12pt,preprint]{aastex}

\slugcomment{}

\shorttitle{Global Properties of X-Ray Flashes and X-Ray-Rich GRBs}
\shortauthors{Sakamoto et al.}

\begin{document}


\title{Global Characteristics of X-Ray Flashes and X-Ray-Rich 
GRBs Observed by HETE-2}


\author{T. Sakamoto\altaffilmark{1,2},
D. Q. Lamb\altaffilmark{4},
C. Graziani\altaffilmark{4},
T. Q. Donaghy\altaffilmark{4},
M. Suzuki\altaffilmark{1},
G. Ricker\altaffilmark{5},
J-L. Atteia\altaffilmark{6},
N. Kawai\altaffilmark{1,2},
A. Yoshida\altaffilmark{2,7},
Y. Shirasaki\altaffilmark{8},
T. Tamagawa\altaffilmark{2},
K. Torii\altaffilmark{19},
M. Matsuoka\altaffilmark{9},
E. E. Fenimore\altaffilmark{3},
M. Galassi\altaffilmark{3},
J. Doty\altaffilmark{5},
R. Vanderspek\altaffilmark{5},
G. B. Crew\altaffilmark{5},
J. Villasenor\altaffilmark{5},
N. Butler\altaffilmark{5},
G. Prigozhin\altaffilmark{5},
J. G. Jernigan\altaffilmark{11},
C. Barraud\altaffilmark{6},
M. Boer\altaffilmark{12},
J-P. Dezalay\altaffilmark{12},
J-F. Olive\altaffilmark{12},
K. Hurley\altaffilmark{11},
A. Levine\altaffilmark{5},
G. Monnelly\altaffilmark{5},
F. Martel\altaffilmark{5},
E. Morgan\altaffilmark{5},
S. E. Woosley\altaffilmark{13},
T. Cline\altaffilmark{14},
J. Braga\altaffilmark{15},
R. Manchanda\altaffilmark{16},
G. Pizzichini\altaffilmark{17},
K. Takagishi\altaffilmark{18},
and M. Yamauchi\altaffilmark{18}}

\altaffiltext{1}{Department of Physics, Tokyo Institute of Technology,
2-12-1 Ookayama, Meguro-ku, Tokyo 152-8551, Japan}
\altaffiltext{2}{RIKEN (Institute of Physical and Chemical Research),
2-1 Hirosawa, Wako, Saitama 351-0198, Japan}
\altaffiltext{3}{Los Alamos National Laboratory, P.O. Box 1663, Los
Alamos, NM, 87545}
\altaffiltext{4}{Department of Astronomy and Astrophysics, University
of Chicago, IL, 60637}
\altaffiltext{5}{Center for Space Research, Massachusetts Institute
of Technology, 70 Vassar Street, Cambridge, MA, 02139}
\altaffiltext{6}{Laboratoire d'Astrophysique, Observatoire
Midi-Pyre\'ne\'es, 14 Ave. E. Belin, 31400 Toulouse, France}
\altaffiltext{7}{Department of Physics, Aoyama Gakuin University,
Chitosedai 6-16-1, Setagaya-ku, Tokyo 157-8572, Japan}
\altaffiltext{8}{National Astronomical Observatory, Osawa 2-21-1,
Mitaka, Tokyo 181-8588, Japan}
\altaffiltext{9}{Tsukuba Space Center, National Space Development
Agency of Japan, Tsukuba, Ibaraki, 305-8505, Japan}
\altaffiltext{10}{Department of Astronomy, New Mexico State University,
1320 Frenger Mall, Las Cruces, NM, 88003-8001}
\altaffiltext{11}{University of California at Berkeley, Space Sciences
Laboratory, Berkeley, CA, 94720-7450}
\altaffiltext{12}{Centre d'Etude Spatiale des Rayonnements, CNRS/UPS,
B.P.4346, 31028 Toulouse Cedex 4, France}
\altaffiltext{13}{Department of Astronomy and Astrophysics, University
of California at Santa Cruz, 477 Clark Kerr Hall, Santa Curz, CA 95064}
\altaffiltext{14}{NASA Goddard Space Flight Center, Greenbelt, MD, 20771}
\altaffiltext{15}{Instituto Nacional de Pesquisas Espaciais, Avenida
Dos Astronautas 1758, Sa\~o Jose\' dos Campos 12227-010, Brazil}
\altaffiltext{16}{Department of Astronomy and Astrophysics, Tata
Institute of Fundamental Research, Homi Bhabha Road, Mumbai, 400 005, India}
\altaffiltext{17}{IASF/CNR Sezione di Bologna, via Gobetti 101, 40129 
Bologna Italy}
\altaffiltext{18}{Faculty of engineering, Miyazaki University, Gakuen
Kibanadai Nishi, Miyazaki 889-2192, Japan}
\altaffiltext{19}{Department of Earth and Space Science, Graduate School of Science,
Osaka University, 1-1 Machikaneyama-cho, Toyonaka, Osaka, 560-0043, Japan}



\begin{abstract}
We describe and discuss the global properties of 45 gamma-ray bursts
(GRBs) observed by HETE-2 during the first three years of its mission,
focusing on the properties of X-Ray Flashes (XRFs) and X-ray-rich GRBs
(XRRs).  We find that the numbers of XRFs, XRRs, and GRBs are
comparable.  We find that the durations and the sky distributions of
XRFs and XRRs are similar to those of GRBs.  We also find that the
spectral properties of XRFs and XRRs are similar to those of GRBs,
except that the values of the peak energy $\eop$ of the burst spectrum
in $\nu F_\nu$, the peak energy flux $\Fp$, and the energy fluence
$\se$ of XRFs are much smaller -- and those of XRRs are smaller -- than
those of GRBs. Finally, we find that the distributions of all three
kinds of bursts form a continuum in the [$\se$(2-30 keV),$\se$(30-400)
keV]-plane, the [$\se$(2-400 keV), $\ep$]-plane, and the [$\Fp$(50-300 
keV), $\ep$]-plane.  These results provide strong evidence that all
three kinds of bursts arise from the same phenomenon.
\end{abstract}




\section{Introduction}

Gamma-ray bursts (GRBs) whose energy fluence $S_X$ in the X-ray energy
band (2-30 keV) is larger than their energy fluence $S_\gamma$ in the
gamma-ray energy band (30-400 keV) have received increasing attention
over the last few years.  In particular, the Wide Field Camera (WFC) on
\bsax\ detected events that were not detected by the Gamma-Ray Burst
Monitor (GRBM) on the same satellite.  These events have been termed
``X-ray flashes'' (XRFs) \citep{heise2000}.  Events for which the ratio
of the fluence in the X-ray energy band is intermediate between those
for XRFs and GRBs have been termed ``X-ray-rich GRBs''
(XRRs).\footnote{Throughout this paper, we define ``X-ray-rich'' GRBs
(XRRs) and ``X-ray flashes'' (XRFs) as those events for which $\log
[S_X(2-30~{\rm kev})/S_\gamma(30-400~{\rm kev})] > -0.5$ and 0.0,
respectively.}  Understanding the relationship between XRFs, XRRs, and
GRBs may provide a deeper understanding of the prompt emission of GRBs.

\section{Observations}

In this paper, we investigate the global properties of a sample of 
HETE-2 bursts.  We require the bursts in this sample to satisfy the
following criteria: (1) the burst is detected in the WXM, (2) the 
burst is localizable by the WXM, and (3) the signal-to-noise of the WXM
data is sufficient to carry out a spectral analysis of the burst. 
Generally, a joint spectral analysis is carried out for the WXM and
the FREGATE data.

Forty-five bursts observed by HETE-2 between the beginning of the
HETE-2 mission and 2003 September 13 met these criteria, and this is
the sample of bursts that we study.  In this study, we consider three
spectral models:  (1) a power law (PL) model whose two parameters are
the power-law index $\alpha$ and the normalization constant $K_{15}$ of
the spectrum at 15 keV; (2) a power law times exponential (PLE) model
whose three parameters are the power-law index $\alpha$, the cutoff
energy $E_0$, and $K_{15}$; and (3) the Band function \citep{band1993}
whose four parameters are the low-energy power-law index $\alpha$, the 
cutoff energy $E_0$, the high-energy power-law index $\beta$, and
$K_{15}$.  We determine whether the data requests a more complicated
model (e.g., the PLE model instead of the PL model, or the Band
function instead of the PLE model) using the maximum likelihood ratio
test, and require a significance $Q < 10^{-2}$ in order to adopt the
more complicated model.

Table 1 gives some information about the localization and the
WXM time histories of the 45 bursts in the sample.  Table 2 gives the
details of the fits made to the time-averaged spectral data for each of
the bursts, including the class of the burst (e.g., XRF, XRR, GRB) and
the spectral parameters of the best-fit spectral model.  Table 3 gives
the photon number and energy fluence of each burst in the 2-30, 30-400,
and  2-400 keV energy bands, and also energy fluence ratio between 2-30
keV and 30-400 keV.  Table 4 gives the photon number peak flux (1
second) of each burst in 2-30 keV, 30-400 keV, 2-400 keV and 50-300 keV
(BATSE Channels 3 and 4; \cite{paciesas1999}) bands.  

When the WXM photon time- and energy-tagged data (TAG data) are
available, we apply a ``cut'' to the WXM data using only the photons
from the pixels on wires in the X and Y detectors that were illuminated
by the burst and that maximize the signal-to-noise ratio (S/N), in the
same manner as we did for GRB 020531 \citep{lamb2003_grb020531}.  We
used the spectral survey data (PHA data for WXM, and SP data for
FREGATE), when TAG data are not available.  The WXM and FREGATE
detector response matrix has been well-calibrated  using observations
of the Crab nebula (WXM; \citet{shirasaki2003_pasj}, FREGATE;
\citet{olive2003_fregate_drm}).  We use the XSPEC version 11.2.0
software package to do the spectral fits.  Details of instruments are
given in \citet{kawai2003} and \citet{shirasaki2003_pasj} for the WXM,
and in \citet{atteia2003} for the FREGATE.    

The time histories of the bursts, details of the spectral fitting
procedure, and time-resolved spectroscopy of some of the bursts are
given in a companion paper (Sakamoto et al 2004b; see also Lamb et al.
2004b).  Other information about the bursts, including skymaps of the
HETE-2 WXM and SXC localizations; the FREGATE $T_{50}$ and $T_{90}$
durations of the bursts; whether an X-ray, optical, or radio afterglow
was detected; whether a host galaxy has been identified; and the
redshift of the burst can be found in the First HETE-2 Burst Catalog
\citep{vanderspek2004}.

\section{X-ray and $\gamma$-ray Fluences} 

\subsection{Distribution of Ratio of X-ray and $\gamma$-ray Fluences}

\begin{figure}[pt]
\centerline{
\includegraphics[width=9cm,angle=-90]{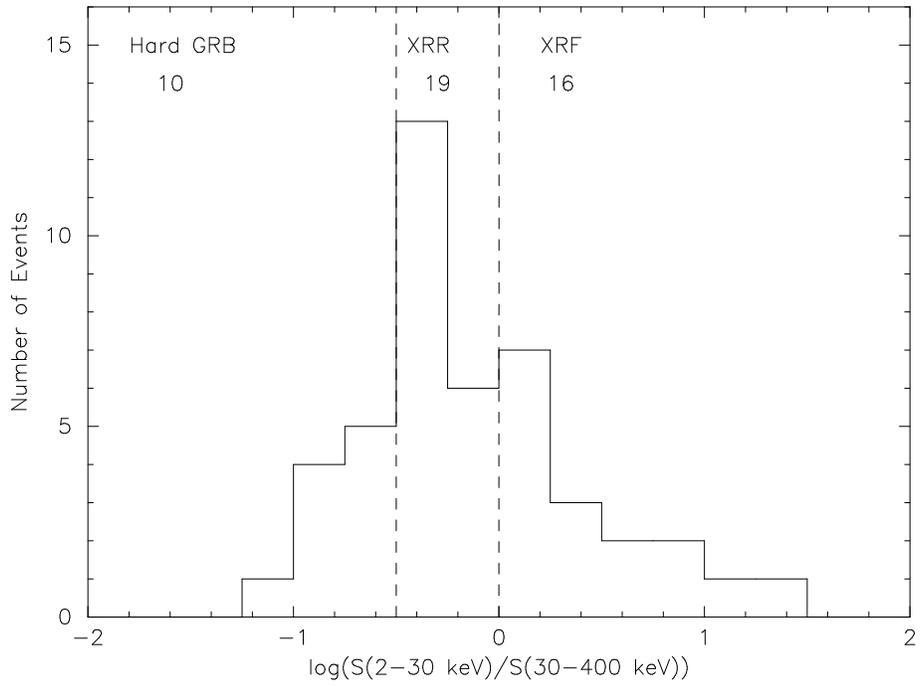}}
\caption{Distribution of the fluence ratio $\se$(2--30
keV)/$\se$(30--400 keV).  The dashed lines correspond to the borders
between hard GRBs and XRRs, and between XRRs and XRFs.}
\label{distri_fluence_ratio}
\end{figure}

The distribution of the fluence ratio $S_{\rm X}$ (2-30
kev)/$S_\gamma$ (30-400 keV) for the 45 bursts in this study is shown
in Figure \ref{distri_fluence_ratio}.   The boundaries between GRBs and
XRRs, and XRRs and XRFs are shown as dashed lines.  The Figure clearly
shows that XRFs, XRRs, and GRBs form a single broad distribution.  The
numbers of XRFs, XRRs, and GRBs, are 16, 19, and 10, respectively.  The
numbers of all three kinds of bursts are roughly equal, modulo the
relatively small sample size. 

\subsection{S$_{X}$ versus S$_{\gamma}$}

\begin{figure}[pt]
\centerline{
\includegraphics[width=9cm,angle=-90]{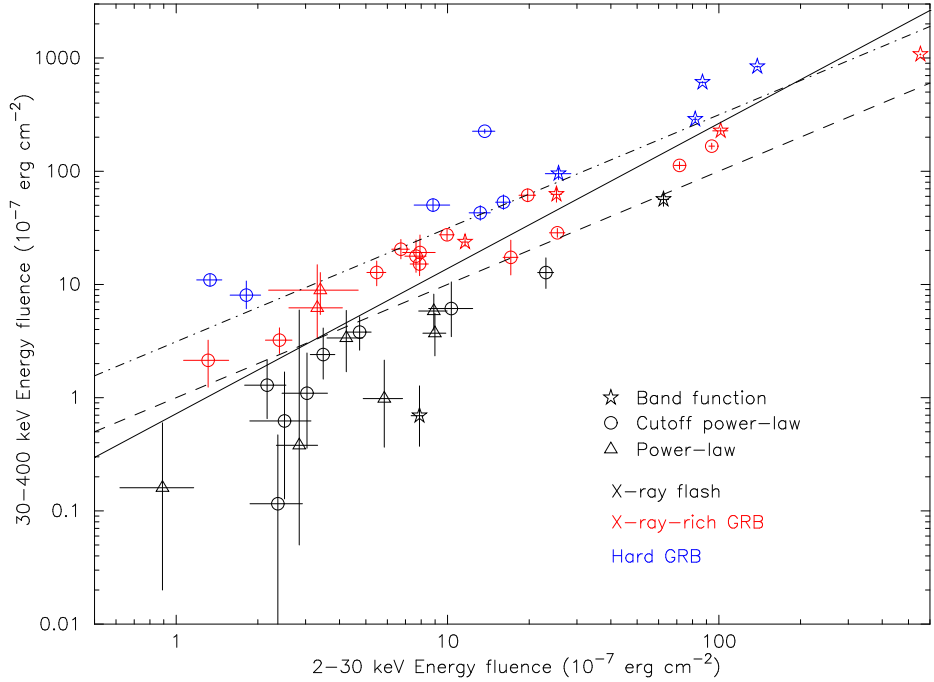}}
\caption{Distribution of the bursts in this study in the [$\se$(2--30
keV,$\se$(30-400 keV]-plane.  The dashed line corresponds to the
boundary between XRFs and XRRs.  The dash-dotted line corresponds to
the boundary between XRRs and GRBs.  The solid line is the best linear
fit to the burst distribution, and is given by $\se$(30--400 keV) =
(0.722$\pm$0.161)  $\times$ $\se$(2--30 keV)$^{1.282 \pm 0.082}$.  The
correlation coefficient of the burst distribution is 0.851.}
\label{s2_30_s30_400}
\end{figure}

Figure \ref{s2_30_s30_400} shows the distributions of XRFs, XRRs, and
GRBs in the [$\se$(2-30 keV), $\se$(30--400 keV)]-plane.  As was
evident in Figure \ref{distri_fluence_ratio}, the three GRB classes
seem to form a single distribution.  There is a strong, tight positive
correlation between $\se$(2-30 keV) and $\se$(30-400 keV): 
$\se$(30-400 keV) = (0.722$\pm$0.161) $\times$ $\se$(2-30 keV)$^{1.282
\pm 0.082}$.  The tightness of the correlation implies that there are
no bursts in the HETE-2 sample with a high X-ray fluence and a low
$\gamma$-ray fluence, or vice versa.  

\section{Durations}
\begin{figure}
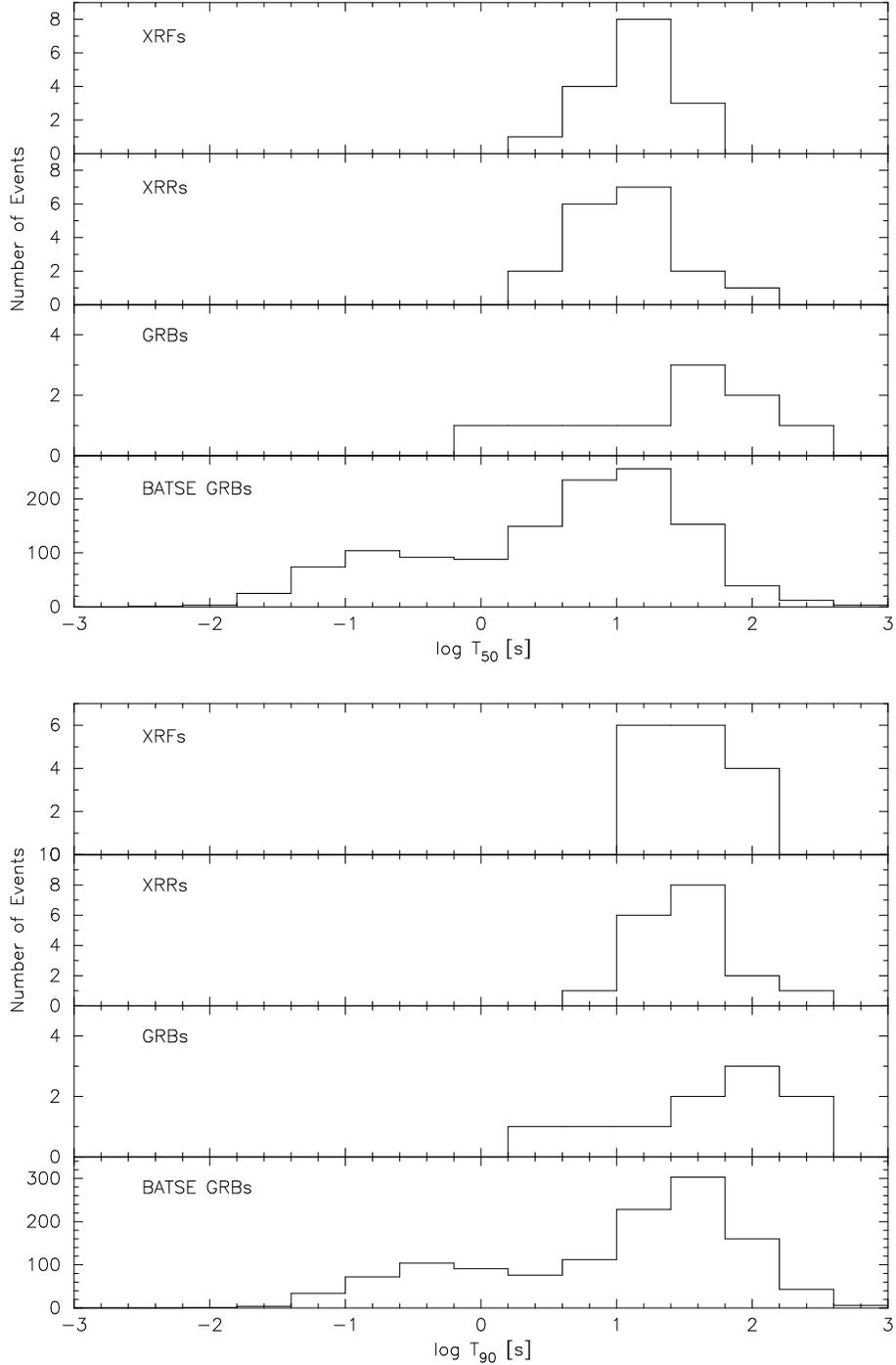

\centerline{
\includegraphics[width=9cm,angle=-90]{t50_hete_grbclass_batse_v2.ps}}
\vspace{0.5cm}
\centerline{
\includegraphics[width=9cm,angle=-90]{t90_hete_grbclass_batse_v2.ps}}
\caption{Comparison between T$_{50}$ (top panel) and T$_{90}$ (bottom
panel) measures of burst duration in the 2--25 keV energy band for the three
kinds of bursts seen by HETE-2 and in the 50-300 keV energy band for
BATSE GRBs.  The subpanels in the top and bottom panels shows (from top
to bottom) the distribution of the durations of XRFs, XRRs, GRBs, and
BATSE GRBs.  The duration of BATSE sample includes not only the long GRBs 
but also the short GRBs.}
\label{t50_t90}
\end{figure}

Figure \ref{t50_t90} shows the distribution of $T_{50}$ (top panel)
and  $T_{90}$ (bottom panel) in the WXM energy band (2-25 keV) for each
kind of GRB.  For comparison, we also show the distribution of
$T_{50}$  and $T_{90}$ for the BATSE bursts \citep{paciesas1999}. 
Although the energy bands in which $T_{50}$ and $T_{90}$ are calculated
are different for HETE-2 and BATSE, the distribution of the durations
of the HETE-2 GRBs are consistent with the distribution of the
durations of the BATSE long GRBs.  There is also no evidence for any
difference in the distribution of durations between the three kinds of
GRBs.  This result is consistent with the {\it Beppo}SAX WFC/CGRO BATSE
sample of XRFs \citep{kippen_xrf_astroph}.  

\section{Sky Distributions}
\begin{figure}
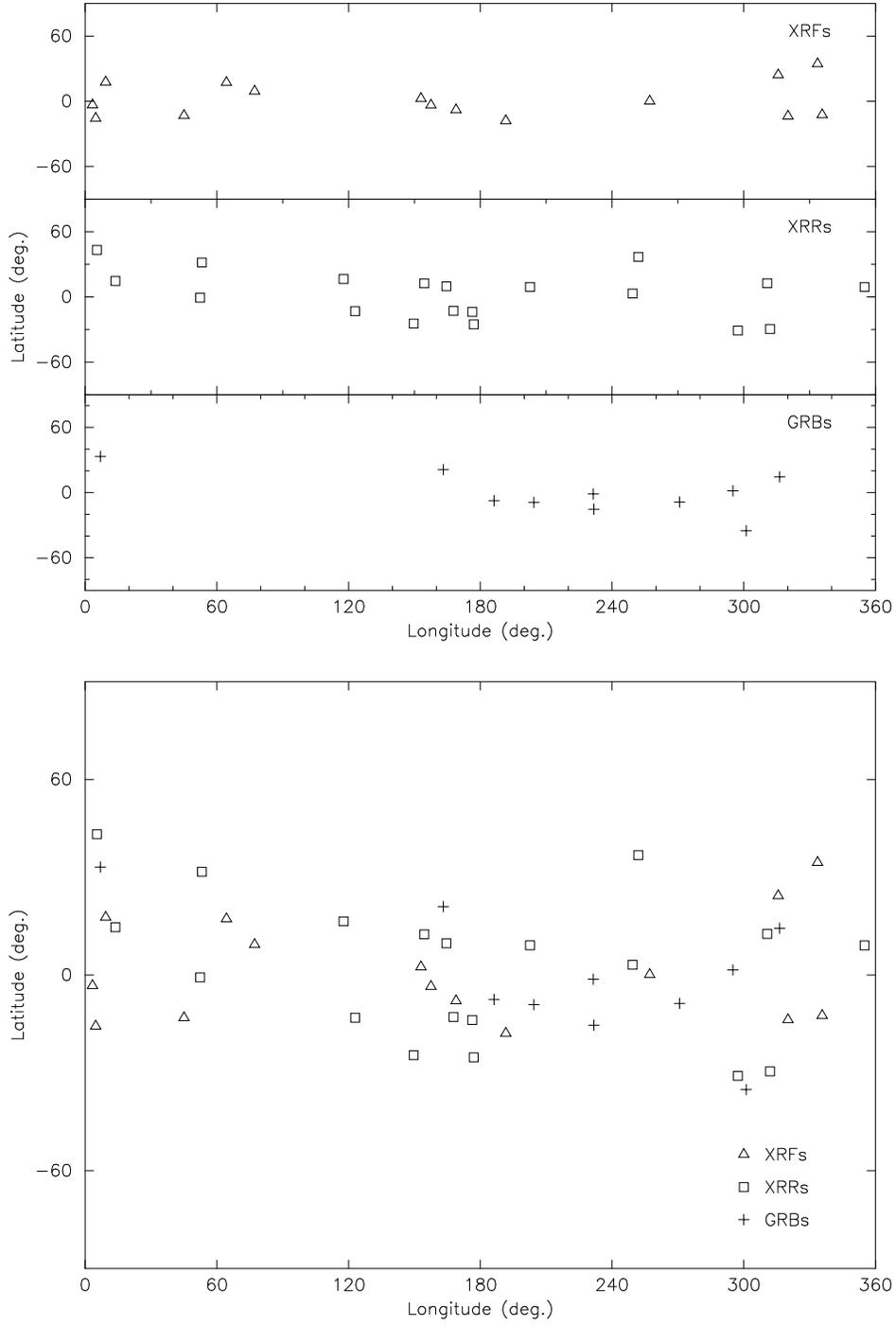

\centerline{
\includegraphics[width=9cm,angle=-90]{3class_eclip_skymap.ps}}
\vspace{0.5cm}
\centerline{
\includegraphics[width=9cm,angle=-90]{all_eclip_skymap.ps}}
\caption{Comparison of the sky distributions in ecliptic coordinates of
the HETE-2 XRFs, XRRs, and GRBs (top three panels), and for all of the
bursts in this study (bottom panel).} 
\label{sky_distributions}
\end{figure}

Figure \ref{sky_distributions} shows the sky distribution in ecliptic
coordinates of HETE-2 XRFs, XRRs, and GRBs (upper three panels), and of
all of the 44 HETE-2 bursts \footnote{Since the attitude control camera 
was not operational, the celestial coordinates of GRB010225 is not 
available.} in this study (bottom panel).  The HETE-2
sky coverage is not uniform, and as a result, it is difficult to make a
meaningful statement about the sky distributions of these three kinds
of GRBs.  Modulo this and the relatively small sample size of each of
the three kinds of bursts, there is no statistically significant
evidence that the sky distributions of the three kinds of bursts are
different.  

\section{Distribution of Spectral Parameters}

We find that a simple PL model provides an adequate fit to the spectral 
data for eight of the 45 bursts in this study.  Six of these bursts
are XRFs and two are XRRs.  In the case of the five XRFs, the slope
of the power-law index is $<$ $-$2.  We inteprete that 
the spectral data for these bursts do not constrain $\eop$ but the fact
that $\beta$ $< -2$ means that $\eop$ is about 2 keV.  This energy 
is near or below the lower limit of the WXM energy band.  
Therefore, we are observing the high-energy power-law portion 
of their Band spectrum and they are XRFs.  
In the case of the four XRRs, the normalization constants
$K_{15}$ of the spectra are the lowest among all of the XRRs and GRBs.
We therefore interprete the lack of evidence for $\eop$ in these bursts
as due to the low signal-to-noise of their spectra.  
In this case, it 
is difficult to constrain the break energy, $E_{0}$, of the spectra and the best 
representable spectral model will be a simple power-law.  

We find that the PLE model provides an adequate fit to the spectral data
for 28 of the 45 bursts in this study.  Eight of these bursts are
XRFs, thirteen are XRRs, and six are GRBs.  The remaining ten bursts
in this study are adequately fit by the Band model but not by any
simpler model.

We do not include two GRBs (GRB020813 and GRB030519) with $\beta > -2$ 
in this study, because they do not represent actual ``peak'' energy 
in $\nu F_{\nu}$ spectrum.  

\subsection{Distribution of $\alpha$-Values}

\begin{figure}[pt]
\centerline{
\includegraphics[width=12cm,angle=0]{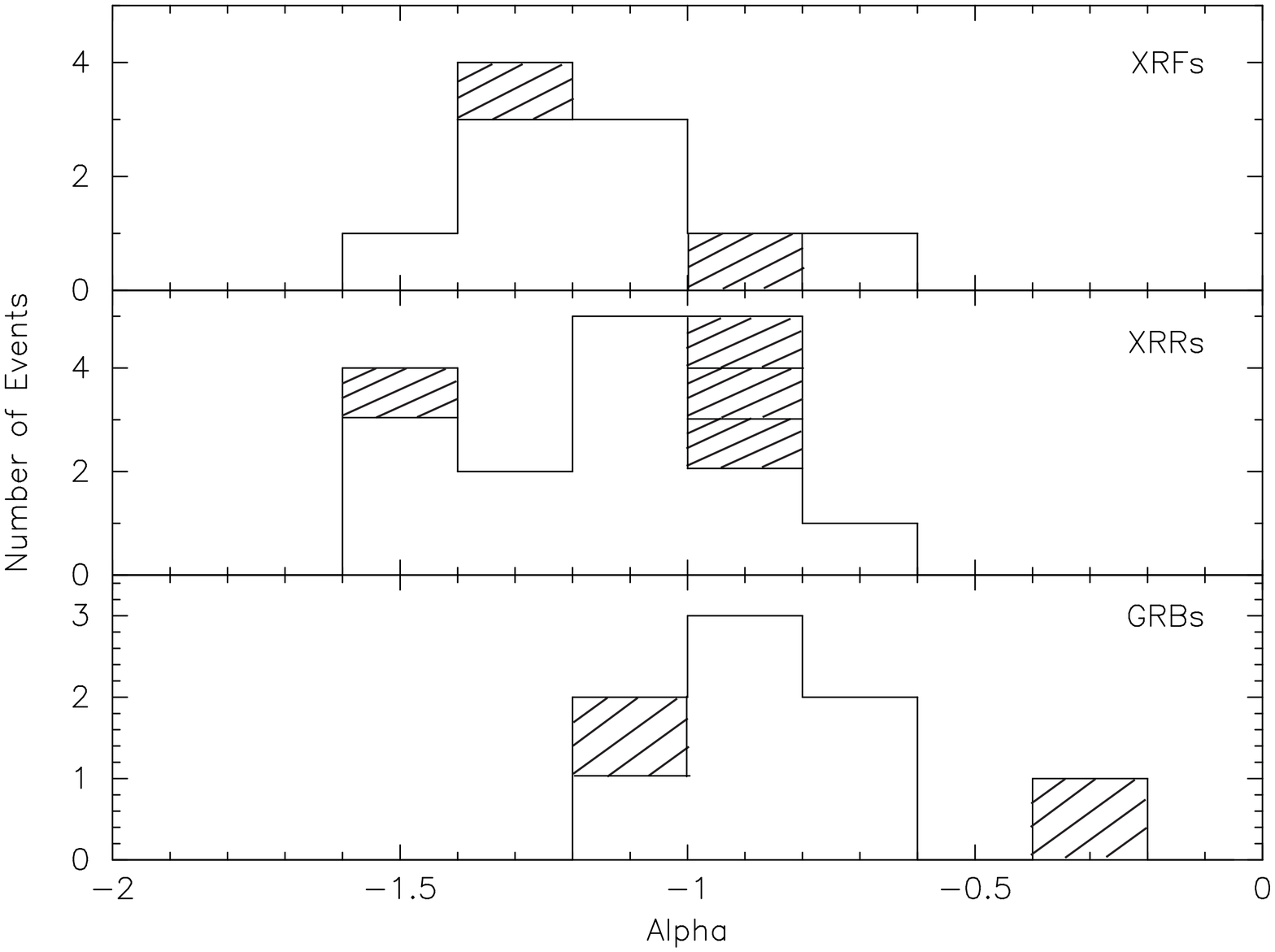}}
\vspace{0.5cm}
\centerline{
\includegraphics[width=9cm,angle=-90]{alpha_all_v2.ps}}
\caption{Distribution of the low-energy power-law index $\alpha$ for each 
of the three kinds of bursts (top panel) and for all of the bursts
(bottom panel).  The hatched $\alpha$-values are the burst speectra 
requiring the Band model and the non-hatched $\alpha$-values are the 
burst spectra that are adequately fit by the PLE model (top panel).}
\label{distri_alpha}
\end{figure}

Figure \ref{distri_alpha} shows the distribution of the low-energy 
photon index $\alpha$.  We include in this figure bursts which require
the PLE model or the Band model in order to adequately represent
their energy spectra.  We do not include bursts whose spectra are
adequately represented by a simple PL model, since in this case the
photon index of the PL model is most likely the high-energy photon
index $\beta$ of the Band model.  
There is a well-known systematic effect when fitting the PLE model to a
spectrum whose shape is that of the Band model but for which the energy
range or the signal-to-noise   of the observations is insufficent to
require the Band model:  the low-energy power-law index $\alpha$ is
smaller (more negative) than it would otherwise be, and the peak energy
$\eop$ of the spectrum in $\nu F_\nu$ is larger than it would otherwise
be.  This systematic effect must be kept in mind when comparing bursts
for which the PLE model adequately represents the data and bursts for
which the Band model is required to adequately represent the data.  We
therefore show as hatched the $\alpha$-values for burst spectra
requiring the Band model and as non-hatched the $\alpha$-values for
burst spectra that are adequately fit by the PLE model.  However, there
is no clear evidence in Figure \ref{distri_alpha} of the above
systematic effect.

The distribution of the low-energy photon index $\alpha$ clusters
around $-1$, and is similar to the BATSE distribution of $\alpha$
values \citep{preece2000}.  The relatively small number of bursts with
$\alpha > -0.5$ in the HETE-2 burst sample compared to the BATSE sample
of  bright bursts \citep{preece2000} could be due to three reasons: 
(1) the HETE-2 burst sample might be lacking very hard GRBs because
such bursts are relatively more difficult for the WXM to detect and to
localize, (2) the HETE-2 values are for time-averaged burst spectra
whereas the $\alpha$ values reported for the BATSE sample of bright
bursts by \citep{preece2000} are for time-resolved spectra; and (3) the
PLE model provides an adequate fit to the spectra of most of the HETE-2
bursts, and therefore the value of $\alpha$ is systematically 
more negative than it would otherwise be, as mentioned above.  The
first reason is unlikely because very hard GRBs are also very intense
(i.e., they have large peak fluxes and fluences).  The second reason
may play a role, since it is well known that the spectra of most bursts
are hardest at or near the peak of the burst time history and softer
afterward.  We regard the third reason as the most likely, since the
vast majority of the 5000 time-resolved burst spectra in the BATSE
sample required the Band model in order to adequately fit the spectrum.

There are no statistically significant differences between the
distributions of $\alpha$ values for XRFs, XRRs, and GRBs (see the top
three panels of Figure \ref{distri_alpha}), although comparison of the
three distributions suffers from small number statistics and from the
presence of the above systematic effect.  Nevertheless, we conclude
that there is no evidence that the distribution of $\alpha$-values for
XRFs, XRRs, and GRBs are different.

\subsection{Distribution of $\eop$-Values}

\begin{figure}[pt]
\centerline{
\includegraphics[width=12cm,angle=0]{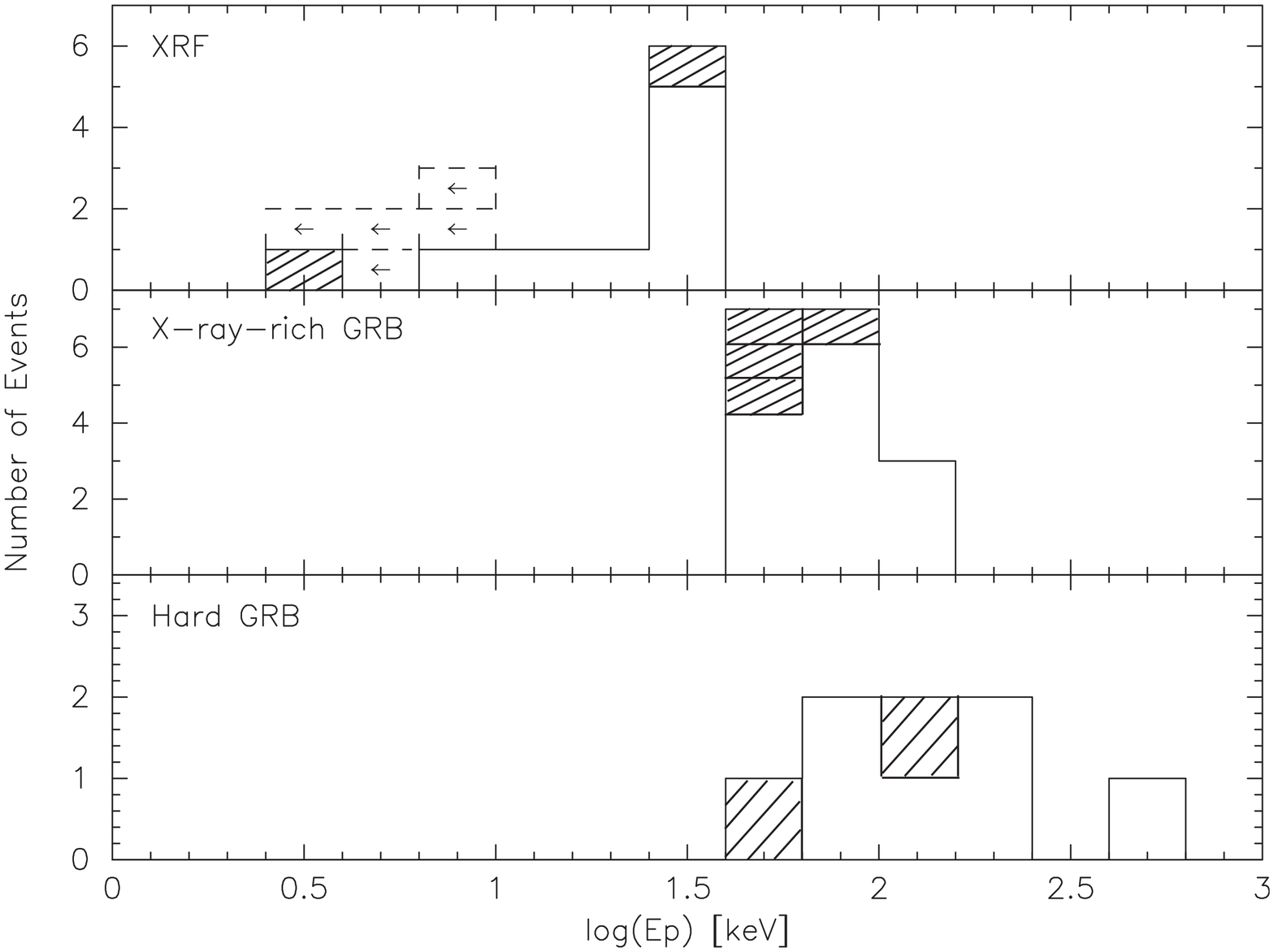}}
\vspace{0.5cm}
\centerline{
\includegraphics[width=9cm,angle=-90]{ep_all_log_v7.ps}}
\caption{Distribution of $E_{\rm peak}$ for each of the three kinds of
bursts (top panel) and for all of the bursts (bottom panel).  
The hatched $\ep$-values are the burst speectra 
requiring the Band model and the non-hatched $\ep$-values are the 
burst spectra that are adequately fit by the PLE model (top panel).}
\label{distri_ep}
\end{figure}

Figure \ref{distri_ep} shows the distribution of the observed peak
energy  $\eop$ of the burst spectra in $\nu F_\nu$.  The events labeled
with left-pointing arrows are the 99.7\% upper limits for $\eop$
derived    using the {\it constrained} Band function
\citep{sakamoto2004a}.  The distribution of $\eop$ is clearly distorted by the
systematic effect mentioned above; i.e., bursts for which the
PLE model provides an adequate representation of the data have values
of $\eop$ that are larger than they would otherwise be.  Despite this
systematic effect, the distribution of $\eop$ values for the sample of
HETE-2 GRBs is much broader than that for the BATSE sample of
time-resolved spectra of bright bursts \citep{preece2000}.  In
particular, the distribution of $\eop$ values in the HETE-2 burst
sample extends to much lower energies.  There are clear differences
between the $\eop$ distributions for XRFs, XRRs, and GRBs, but this is
simply because of the strong correlation that must exist between $\eop$
and the fluence ratio $\se$(2-30 keV)/$\se$(30-400 keV).  This is the fact
that we are classifying the GRBs for  30 keV as a boundary.  The $\ep$
distributions of the XRRs and the hard GRBs are quite similar.  

\begin{figure}
\centerline{
\includegraphics[width=9cm,angle=-90]{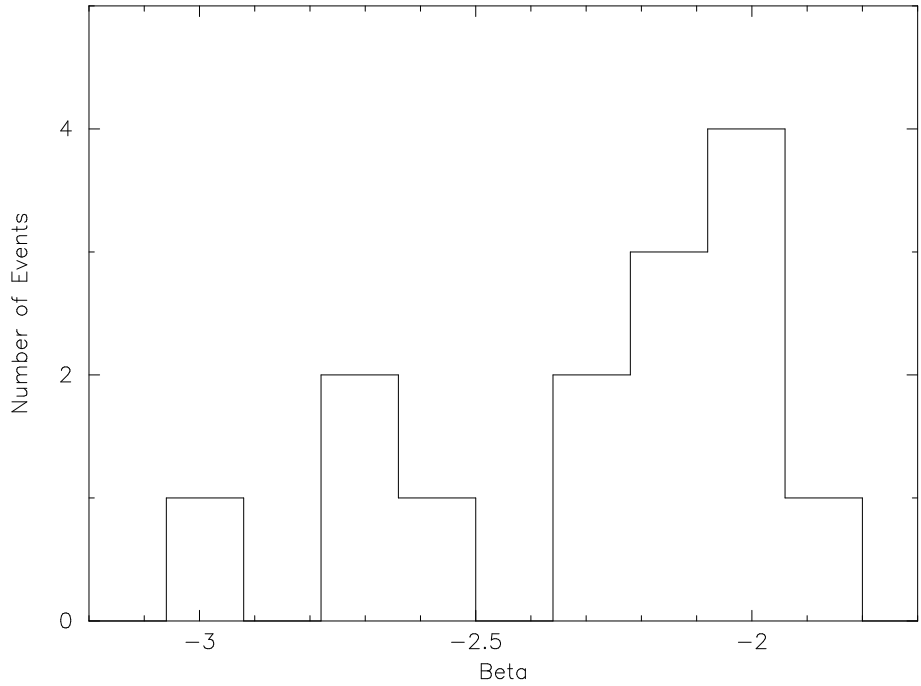}}
\caption{Distribution of the high-energy power-law index $\beta$ for all of
the bursts for which $\beta$ could be determined.  Two GRBs (GRB020813 and 
GRB030519) with $\beta > -2$ are not included in the sample, because they 
do not represent actual ``peak'' energy in $\nu$F$_{\nu}$ spectrum.}
\label{distri_beta}
\end{figure}

The distribution of $\beta$ is shown in Figure \ref{distri_beta}.  
Because of the small number of GRBs with significantly constrained
$\beta$,  only the distribution for all the GRB classes are plotted. 
The  distribution of $\beta$ is similar to the BATSE GRBs 
\citep{preece2000}.

\section{Correlations Between $\eop$ and Other Burst Properties}

\subsection{$\eop$ vs. Fluence Ratio}

\begin{figure}[pt]
\centerline{
\includegraphics[width=9cm,angle=-90]{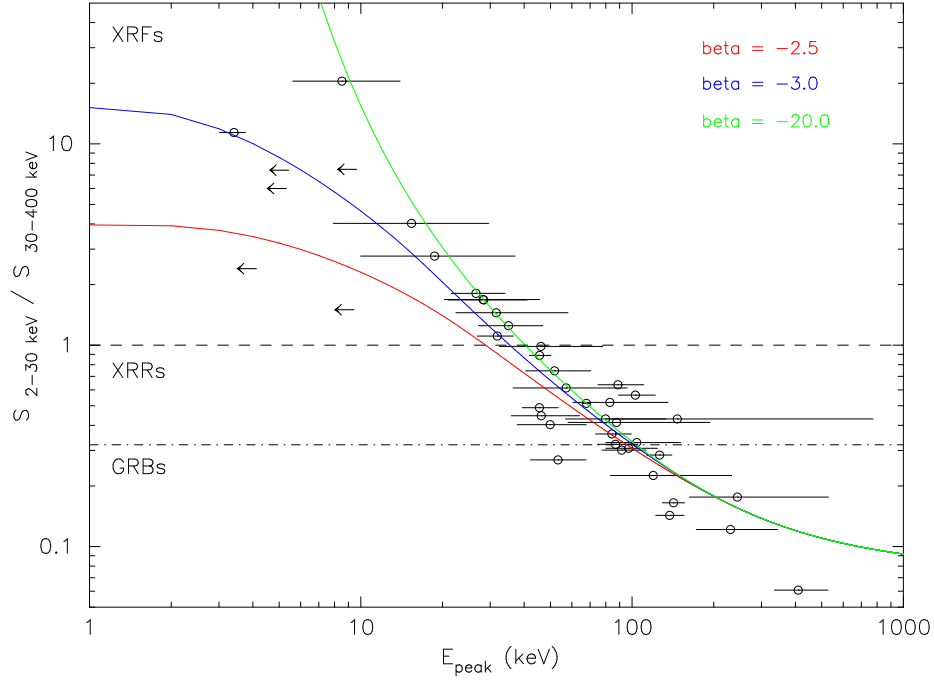}}
\caption{Distribution of bursts in the [$E_{\rm peak}$,$\se$(2--30
keV)/$\se$(30--400 keV]-plane.  Overlaid are curves corresponding to
the X-ray to $\gamma$-ray fluence ratio as a function of $\eop$,
assuming the Band function with $\alpha$ = $-$1 and $\beta$ = $-$2.5
(red), $-$3.0 (blue), and $-$20.0 (green).}
\label{ep_fluence_ratio}
\end{figure}

Figure \ref{ep_fluence_ratio} shows the distribution of observed peak
energy $\eop$ versus the fluence ratio $\se$(2-30 keV)/$\se$(30-400
keV).   Since the fluence ratio is independent of the normalization
parameter of the model spectrum, it is possible to calculate the
relationship between the fluence ratio and $\eop$.  The overlaid
curves in Figure \ref{ep_fluence_ratio} are the calculated
relationships, assuming the Band function, for $\alpha$ = $-$1 and
$\beta$ = $-$2.5 (red), $-$3.0 (blue), and  $-$20.0
(green).  The dependence of the fluence ratio on $\beta$ is weak when
$\eop$ is greater than 30 keV, and understandably, becomes strong when
$\eop$ is less than 30 keV.  This implies that the choice of the proper
spectral model is important for determining the fluence ratio, and for
determining which bursts are XRFs and XRRs.  Fortunately, the
importance of choosing the correct spectral model for the latter is
modest because a range in $\beta$ of ($-$2)-($-$20) produces a range in the
fluence ratio of only 40\% at $\eop$ = 30 keV, which corresponds to the
boundary between XRFs and XRRs.

\subsection{$\alpha$ and $\beta$ vs. $\eop$}

\begin{figure}[pt]
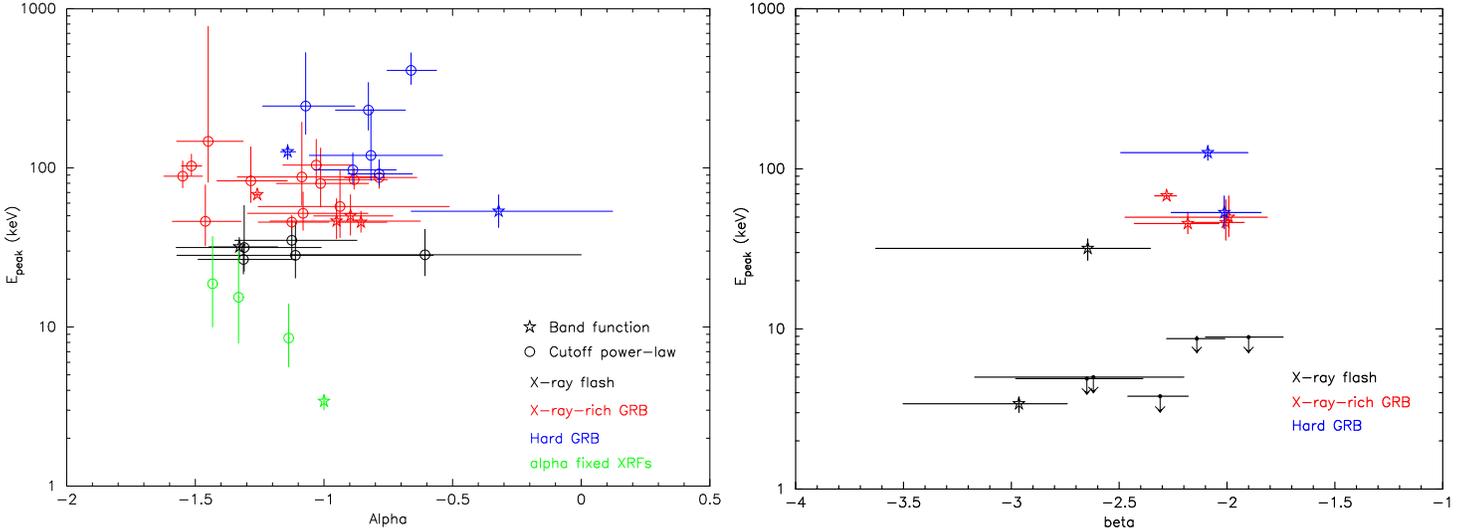

\centerline{
\includegraphics[width=7cm,angle=-90]{alpha_ep_v2.ps}
\includegraphics[width=7cm,angle=-90]{beta_ep_v2.ps}}
\caption{The low-energy power-law index $\alpha$ (left panel) and
$\beta$ (right panel) vs. $\eop$.  The three kinds of bursts are
denoted by different colors (XRF: black; XRR: red; and hard GRB:  blue)
and different symbols indicate the different best-fit spectral models 
(circle: PLE model; star: Band function).  Also plotted are 
the XRFs for which the value of $\alpha$ was fixed (green).}
\label{alpha_beta_ep}
\end{figure}

Figure \ref{alpha_beta_ep} shows the distribution of $\alpha$-values
(left panel) and $\beta$-values (right panel) versus $\eop$.  $\alpha$ 
and $\beta$ show no statistically significant correlation with $\eop$, and 
therefore none with the kind of burst. \citet{kippen_xrf_astroph} also
found no statistically significant correlation between $\alpha$ and
$\eop$ in the {\it Beppo}SAX WFC/CGRO BATSE sample of XRFs and GRBs.

\subsection{2-400 keV Fluence vs. $\eop$}

\begin{figure}[pt]
\centerline{
\includegraphics[width=9cm,angle=-90]{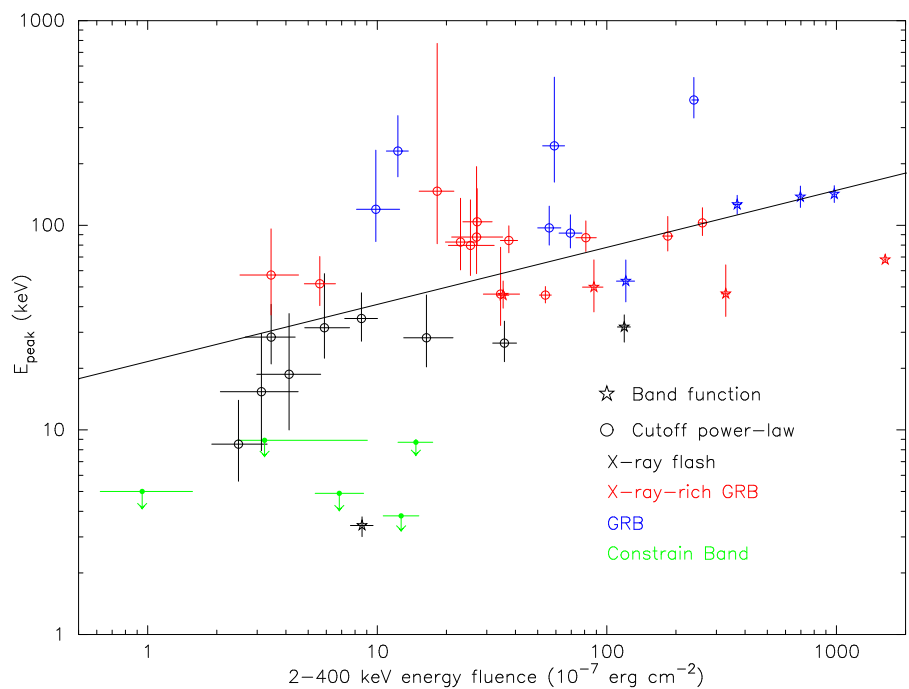}}
\caption{Distribution of bursts in the [$\se$(2--400 keV),$\eop$]-plane.  The solid line is the best
linear fit to the burst distribution, and is given by $\eop$ =
(21.577$\pm$4.656) $\times$ [$\se$(2--400 keV)/10$^{-7}$ ergs cm$^{-2}$
s$^{-1}$] $^{0.279\pm0.053}$.  The correlation coefficient for the
burst distribution is 0.511.}  \label{fluence2_400kev_ep}
\end{figure}

The correlation between the fluence in 2--400 keV and $\eop$ are shown
in Figure \ref{fluence2_400kev_ep}.  This figure shows the correlation between
$\se$(2--400 keV) and $\eop$.  The best-fit power-law slope between
$\eop$ and $\se$(2--400 keV) is 0.279 $\pm$ 0.053.  Thus, while the
scatter in the correlation is large (the correlation coefficient is
0.511), the significance of the correlation is also large.

\subsection{Peak Photon Number Flux vs. $\eop$}

\begin{figure}[pt]
\centerline{
\includegraphics[width=9cm,angle=-90]{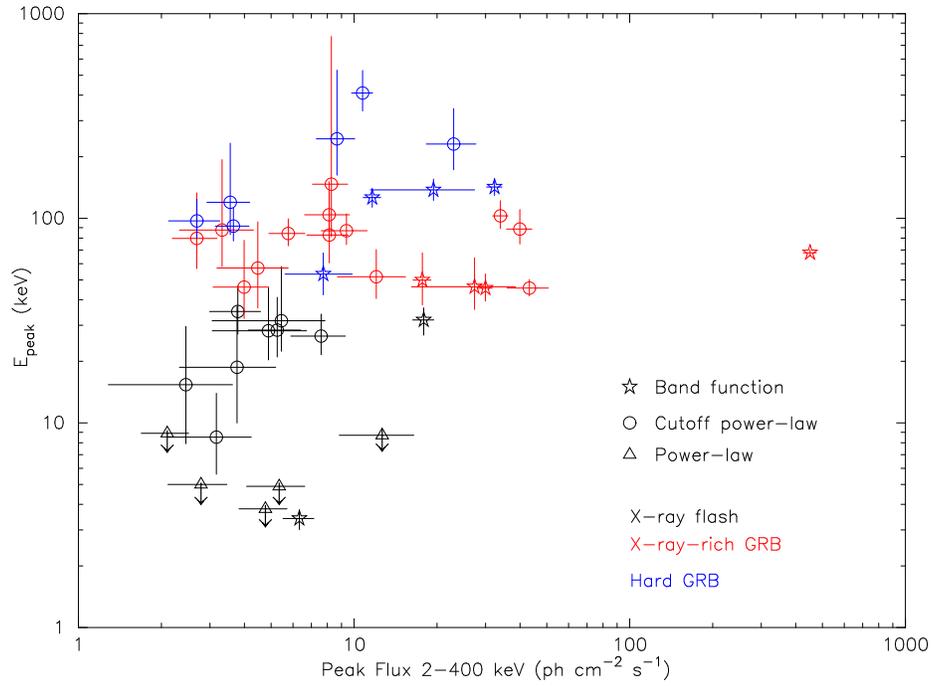}}
\caption{Distribution of bursts in the [$\fn$(2--400 keV),$\eop$]-plane.  The 
correlation coefficient for the burst distribution is 0.297.} 
\label{pf2_400kev_ep}
\end{figure}

\begin{figure}[pt]
\centerline{
\includegraphics[width=9cm,angle=-90]{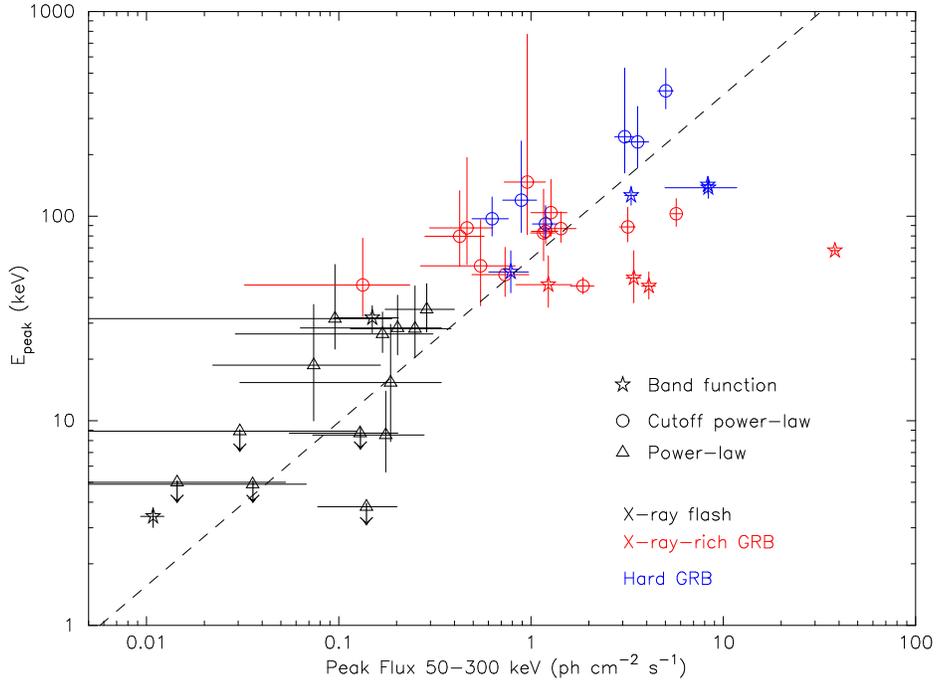}}
\caption{Distribution of bursts in the [$\fn$(50--300 keV),$\eop$]-plane. 
The dashed line corresponds to the best linear fit to the burst
distribution and is given by $\eop$ = 62.02$\pm$1.71
$\fn$(50--300 keV)$^{0.80\pm0.32}$.  The correlation coefficient for
the burst distribution is 0.802.}
\label{pf50_300_ep}
\end{figure}

Figures \ref{pf2_400kev_ep} and \ref{pf50_300_ep} show the
distribution  of HETE-2 bursts in the [$\fn$(2--400 keV),$\eop$]-plane
and the [$\fn$(50--300 keV),$\eop$]-plane, respectively.  There is no 
evidence for a correlation between $\eop$ and the peak photon flux
$\fn$(2--400 keV), while a strong correlation exists between $\eop$ and
the peak photon flux $\fn$(50--300 keV) (the latter has a correlation
coefficient of 0.802).  \citet{kippen_xrf_astroph} suggested a similar
correlation for the WFC/BATSE sample of XRFs and GRBs.  

However, the correlation between $\eop$ and $\fn$(50-300 keV) is an
artifact of the choice of 50-300 keV for the energy band in which the
peak flux is measured.  The reason is that for GRBs $\fn$(50--300 keV)
is roughly the {\it bolometric} peak photon number flux, whereas for
XRRs, and especially for XRFs, $\fn$(50--300 keV) it is clearly not. 
This is because $\eop$ lies near or below the lower limit of this
energy band for XRRs, and far below the lower limit of the energy band
for XRFs.  The result is that the peak photon number fluxes for these
bursts are greatly reduced from their bolometric values, as can be
clearly seen by comparing Figures \ref{pf2_400kev_ep} and
\ref{pf50_300_ep}.

\begin{figure}
\centerline{
\includegraphics[width=9cm,angle=-90]{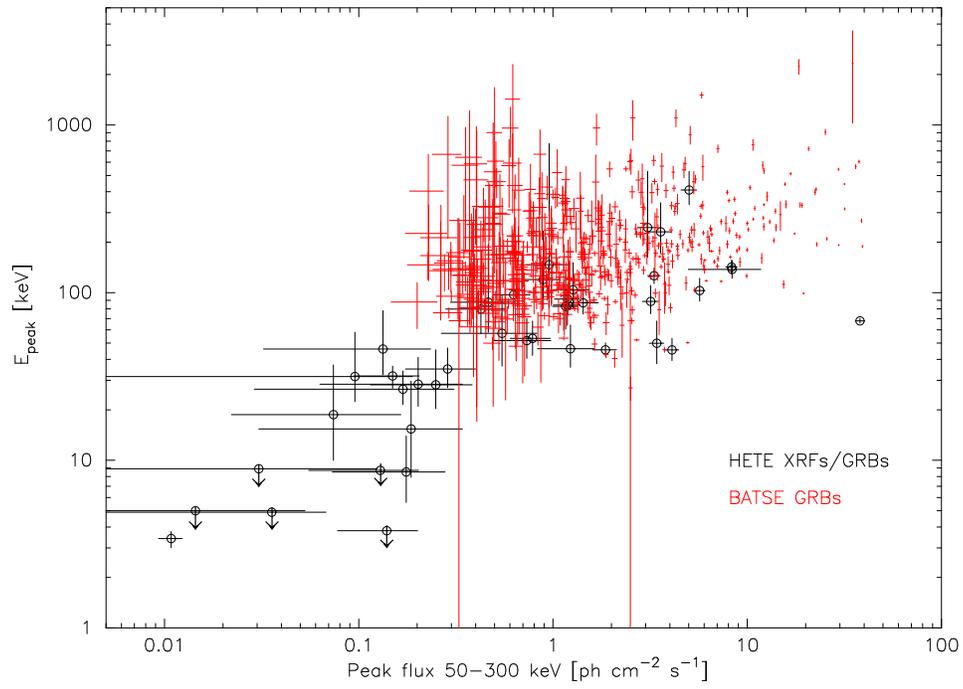}}
\caption{Distribution of HETE-2 bursts (black) and BATSE bursts (red) in
 the [$\fn$(50--300 keV),$\eop$]-plane.}
\label{batse_grb_wfc_batse_xrf_hete}
\end{figure}

\begin{figure}
\centerline{
\includegraphics[width=9cm,angle=-90]{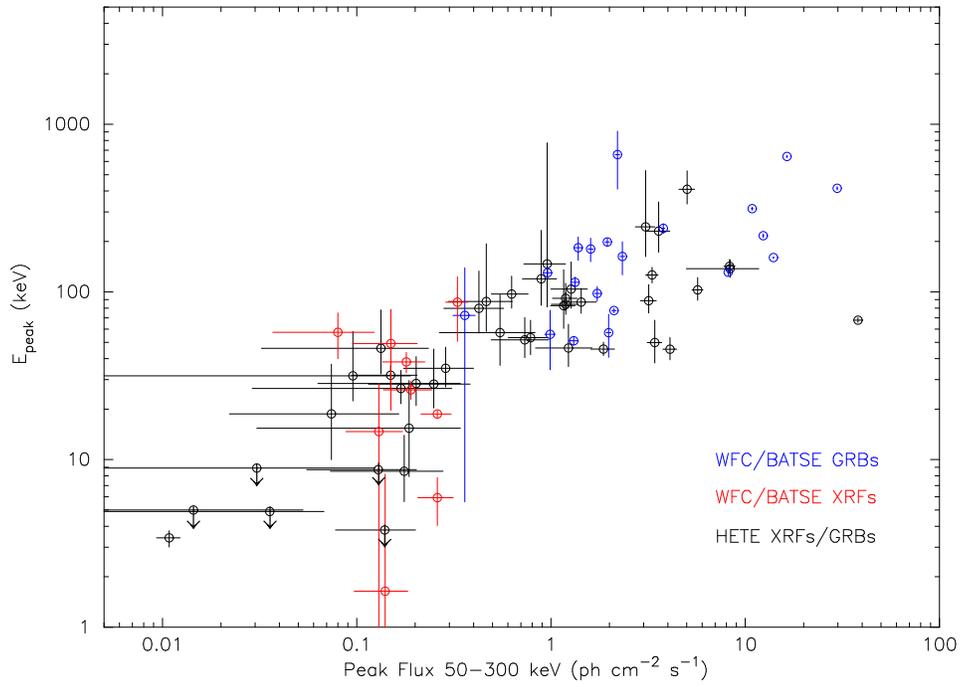}}
\caption{Distribution of HETE-2 bursts (black) and WFC/BATSE bursts (red
and blue) in the [$\fn$(50--300 keV),$\eop$]-plane.}
\label{wfc_batse_hete_xrf}
\end{figure}

Figures \ref{batse_grb_wfc_batse_xrf_hete} and \ref{wfc_batse_hete_xrf}
compare the distribution of HETE-2 bursts in the [$\fn$(50-300
keV),$\eop$]-plane with the distribution of BATSE bursts and the
distribution of WFC/BATSE bursts, respectively, in the same plane.  The
distribution of HETE-2 bursts is consistent with the distribution of
BATSE bursts for $\eop >$ 50 keV but extends farther down in $\eop$
(and therefore in $\fn$(50-300 keV)).  This is expected because of the
BATSE trigger threshhold, which is 50 keV.  The distribution of HETE-2
bursts is consistent with the distribution of WFC/BATSE bursts but also
extends down to fainter peak photon number fluxes for a similar reason.

\section{Discussion}

\subsection{Comparison of XRF, XRR, and GRB Properties}

\begin{figure}
\centerline{
\includegraphics[width=9cm,angle=-90]{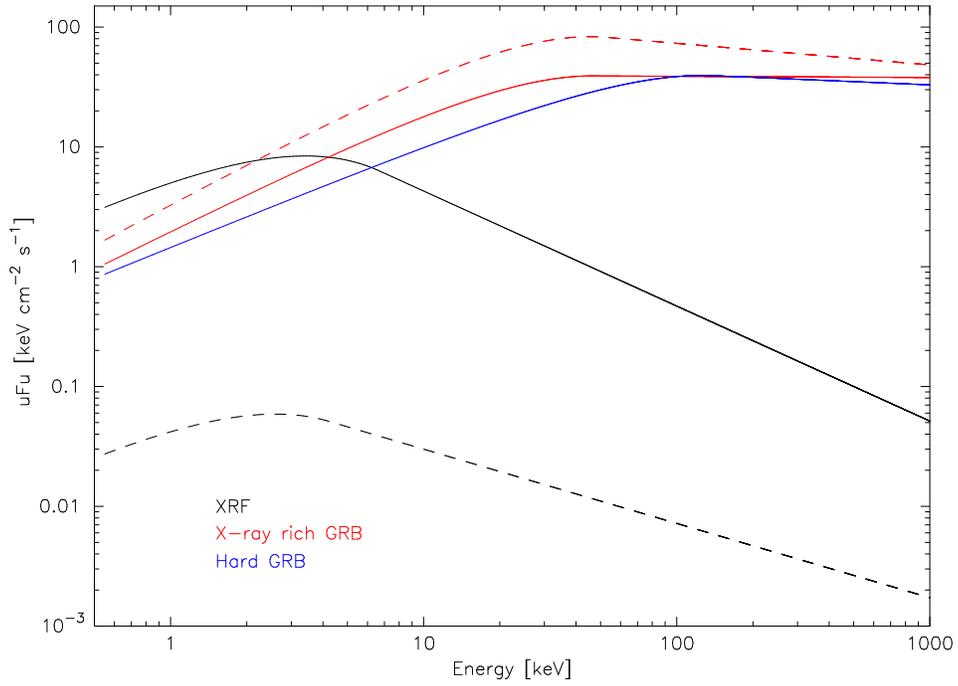}}
\caption{Examples of best-fit $\nu F_{\nu}$ spectra for XRFs (black) 
GRB010213 (solid) and GRB020903 (dash), XRRs (red) GRB010613 (solid) 
and GRB021211 (dash), and GRBs (blue) GRB030328 (solid).}
\label{spectral_shape}
\end{figure}

We have studied the global properties of 45 GRBs localized by the
HETE-2 WXM during the first three years of its mission, focusing on the
properties of XRFs and XRRs.  We find that the numbers of XRFs, XRRs,
and GRBs are comparable for bursts localized by the HETE-2 WXM.  We
find that there is no statistically significant evidence for any
difference in the duration distributions or the sky distributions of
the three kinds of bursts.  We also find that the spectral properties
of XRFs and XRRs are similar to those of GRBs, except that the values
of the peak energy $\eop$ of the burst spectrum in $\nu F_\nu$, the
peak flux $\Fp$ ,and the fluence $\se$ of XRFs are much smaller -- and
those of XRRs are smaller -- than those of GRBs.  
Our results are consistent with \citet{celine2003} who studied the 
spectral properties of the HETE-2 GRBs using the FREGATE data.  
Figure
\ref{spectral_shape}, which shows the best-fit $\nu F_\nu$ spectra of
two XRFs, two XRRs, and two GRBs, illustrates this.  Finally, we find
that the distributions of all three kinds of bursts form a continuum in
the [$S$(2-30 keV), $S$(30-400 keV)]-plane, the [$S$(2-400 keV),
$\ep$]-plan, and the [$F_{\rm peak}$(50-300  keV), $\eop$]-plane.  These
results provide strong evidence that all three kinds of bursts arise
from the same phenomenon.

\subsection{Theoretical Models of XRFs}

Several theoretical models of XRFs have been proposed.  GRBs at very
high redshifts might be observed as XRFs \citep{heise2000}.  However, the
fact that the duration distribution for XRFs is similar to that for
GRBs argues against this hypothesis as the explanation of most XRFs, as
does the low redshifts \citep{soderberg2004,fynbo2004} and the
redshift constraints \citep{bloom2003} that exist for several XRFs. 

According to \citet{meszaros2002} and \citet{woosley2003}, X-ray
(20-100 keV) photons are produced effectively by the hot cocoon
surrounding the GRB jet as it breaks out, and could produce XRF-like
events if viewed well off the axis of the jet.  However, it is not
clear that such a model would produce roughly equal numbers of XRFs,
XRRs, and GRBs, or the nonthermal spectra exhibited by XRFs.

\cite{yamazaki2002,yamazaki2003} have proposed that XRFs are the result
of a highly collimated GRB jet viewed well off the axis of the jet.  In
this model, the low values of $\ep$ and $\eiso$ (and therefore for
$\eop$ and $\se$) seen in XRFs is the result of relativistic beaming. 
However, it is not clear that such a model can produce roughly equal
numbers of XRFs, XRRs, and GRBs, and still satisfy the observed
relation between $\eiso$ and $\ep$ \citep{amati2002,lamb2004_ep_eiso}.
 
The ``dirty fireball'' model of XRFs posits that baryonic material is
entrained in the GRB jet, resulting in a bulk Lorentz factor $\Gamma$
$\ll$  300 \citep{dermer1999,huang2002,dermer2003}.  At the opposite
extreme, GRB jets in which the bulk Lorentz factor $\Gamma$ $\gg$ 300
and the contrast between the bulk Lorentz factors of the colliding
relativistic shells in the internal shock model are small can also
produce XRF-like events \citep{mochkovitch2003}.

It has been proposed that XRFs are due to universal GRB jets in which
the luminosity falls off like a power law from the jet axis
\citep{zhang2002,rossi2002} and are viewed well off the jet axis
\citep{zhang2004}.  However, \cite{lamb2004} have shown that such a
model predicts far more XRFs than GRBs, in conflict with the HETE-2
results described in this paper.  A universal GRB jet model in which
the luminosity falls off like a Gaussian may do better
\citep{zhang2004}.

\cite{lamb2004} have shown that a unified description of XRFs, XRRs,
and GRBs is possible in a model in which the GRB jet opening angle
varies over a wide range.  In this model, XRFs are due to jets with
wide opening angles while GRBs are due to jets with narrow opening
angles.

As this discussion suggests, understanding the  properties of XRFs and
XRRs, and clarifying the relationship between these two kinds of events
and GRBs, could provide a deeper understanding of the prompt emission
of GRBs.  And as \cite{lamb2004} have emphasized, XRFs may provide
unique insights into the nature of GRB jets, the rate of GRBs, and the
relationship between GRBs and Type Ic supernovae.

\section{Conclusions}

We have studied the global properties of 45 GRBs observed by HETE-2
during the first three years of its mission, focusing on the properties
of XRFs and XRRs.  We find that the numbers of XRFs, XRRs, and GRBs are
comparable.  We find that the durations and the sky distributions of
XRFs and XRRs are similar to those of GRBs.  We also find that the
spectral properties of XRFs and XRRs are similar to those of GRBs,
except that the values of the peak energy $\eop$ of the burst spectrum
in $\nu F_\nu$, the peak flux $\Fp$ ,and the fluence $\se$ of XRFs are
much smaller -- and those of XRRs are smaller -- than those of GRBs. 
Finally, we find that the distributions of all three kinds of bursts
form a continuum in the [$S$(2-30 keV), $S$(30-400 keV)]-plane, the
[$S$(2-400 keV), $\ep$]-plan, and the [$F_{\rm peak}$(50-300  keV),
$\eop$]-plane.   These results provide strong evidence that all three
kinds of bursts arise from the same phenomenon.  They also provide
constraints on theoretical models of XRFs.

\acknowledgments

We would like to thank Dr. R. Marc Kippen for providing us the WFC/BATSE 
spectral parameters.  
The {\it HETE} mission is supported in the U.S. by NASA contract NASW-4690;
in Japan, in part by the Ministry of Education, Culture, Sports,
Science, and Technology Grant-in-Aid 12440063; and in France, by CNES
contract 793-01-8479.  K. Hurley is grateful for {\it Ulysses} support
under contract JPL 958059 and for {\it HETE} support under contract
MIT-SC-R-293291.  G. Pizzichini acknowledges support by the Italian Space
Agency.



\begin{deluxetable}{cccccccccccccc}
\tabletypesize{\scriptsize}
\tablecaption{Some Properties of 45 HETE-2 GRBs\label{hyo:hete_grb_summary}}
\tablewidth{0pt}
\tablehead{
\colhead{GRB} & \colhead{BID} & \colhead{$\theta_{X}$} &
\colhead{$\theta_{Y}$} & \colhead{TT$^{a}$} & \colhead{TS$^{b}$} &
\colhead{EB$^{c}$} & \colhead{R.A.} & \colhead{Dec.} & \colhead{$l$} &
\colhead{$b$} &\colhead{error$^{d}$}
& \colhead{t$_{50}$ (WXM)} & \colhead{t$_{90}$ (WXM)}}
\rotate
\startdata
GRB010213   & 10805 &  -2.41 &  13.60      & --- & --- & --- &
10h31m36s & +05d30m39s & 239.6 & 50.3 & 30.2$^{\prime}$ & 8.6 $\pm$ 1.2 & 24.5
$\pm$ 1.2\\
GRB010225$^{e}$   & 1491  & -23.10 &   0.97      & G & 1.3s & 5-120 & --- &
--- & --- & --- & --- & 6.2 $\pm$ 1.3 & 15.9 $\pm$ 3.9\\
GRB010326B  & 1496  &   7.97 & -15.02 & G & 160ms & 5-120 & 11h24m24s
& -11d09m57s &  271.2 & 46.3 & 36$^{\prime}$ & 1.7 $\pm$ 0.2 & 5.2 $\pm$ 0.2\\
GRB010612   & 1546  &  13.81 &   1.17 & G & 160ms & 30-400 & 18h03m18s
& -32d08m01s & 359.2 &  -4.9 & 36$^{\prime}$ & 17.4 $\pm$ 0.8 & 28.5 $\pm$
0.2\\
GRB010613   & 1547  & -30.50 &  25.17 & G & 1.3s & 30-400 & 17h00m40s
& +14d16m05s &  33.9 & 30.9 & 36$^{\prime}$ & 23.8 $\pm$ 1.2 & 51.8 $\pm$ 0.7\\
GRB010629B  & 1573  & -26.60 &   8.29 & G & 1.3s & 5-120 & 16h32m38s &
-18d43m24s & 358.6 & 19.5 & 15$^{\prime}$ & 9.3 $\pm$ 0.3 & 16.2 $\pm$ 0.2\\
GRB010921   & 1761  & -23.95 &  39.45 & G & 1.3s & 5-120 & 23h01m53s &
+44d16m12s & 103.1 & -14.3 & 20$^{\circ}$$\times$15$^{\prime}$ & -- & -- \\
GRB010928   & 1770  &  -2.99 &  35.00 & G & 1.3s & 30-400 & 23h28m55s
& +30d39m11s & 102.9 & -26.7 & 16.4$^{\prime}$ $\times$ 11$^{\circ}$
& 29.5 $\pm$ 3.5 & 59.0 $\pm$ 1.8\\
GRB011019   & 10823 & -18.29 & -17.63 & --- & --- & --- & 00h42m50s &
-12d26m58s & 114.7 & -75.2 & 35$^{\prime}$ & 12.2 $\pm$ 1.3 & 31.6 $\pm$ 1.2\\
GRB011103   & 1829  &  -0.32 & -10.94 & XG & 5.12s & --- & 03h20m37s &
17d40m01s & 166.1 & -32.4 & --- & 8.6 $\pm$ 1.7 & 19.7 $\pm$ 1.2\\
GRB011130   & 1864  & -13.03 &  22.83 & XG & 5.12s & --- & 03h05m36s &
+03d48m36s & 174.4 & -45.2 & 10$^{\prime}$ & 23.8 $\pm$ 0.6 & 39.5 $\pm$ 0.4\\
GRB011212   & 10827 &  -1.60 &   9.71 & --- & --- & --- & 05h00m05s &
+32d07m39s & 171.8 & -6.3 & 11$^{\prime}$ & 33.2 $\pm$ 1.2 & 72.5 $\pm$ 2.8\\
GRB020124   & 1896  &  14.65 & -31.57 & G & 1.3s & 30-400 & 09h32m49s
& -11d27m35s & 244.9 & 28.3 & 12$^{\prime}$ & 18.6 $\pm$ 1.1 & 50.2 $\pm$ 2.3\\
GRB020127   & 1902  &  -7.51 &  20.76 & G & 5.12s & 30-400 & 08h15m06s
& +36d44m31s & 184.7 & 31.8 & 8$^{\prime}$ & 6.0 $\pm$ 0.3 & 17.6 $\pm$ 1.9\\
GRB020317   & 1959  & -17.14 &  15.15 & G & 1.3s & 5-120 & 10h23m21s &
+12d44m38s & 228.1 & 52.5 & 18$^{\prime}$ & 2.4 $\pm$ 0.4 & 14.7 $\pm$ 0.5\\
GRB020331   & 1963  &   6.91 & -14.33 & G & 160ms & 30-400 & 13h16m34s
& -17d52m29s & 311.3 & 44.6 & 10$^{\prime}$ & 35.7 $\pm$ 1.8 & 78.7 $\pm$ 1.8\\
GRB020531   & 2042  &  22.94 &  11.33 & G & 20ms & 30-400 & 15h14m45s
& -19d21m35s & 343.6 & 32.0 & 38$^{\prime}$ & 1.1 $\pm$ 0.2 & 2.5 $\pm$ 0.3\\
GRB020625   & 2081  &   5.64 &  10.12 & G & 5.2s & 30-400 & 20h44m14s
& +07d10m12s & 53.3 & -21.1 & 13.8$^{\prime}$ & 13.5 $\pm$ 1.2 & 119.2 $\pm$
2.4\\
GRB020801   & 2177  &   4.73 &  35.44 & G & 1.3s & 30-400 & 21h02m14s
& -53d46m13s & 343.9 & -40.7 & 13.9$^{\prime}$ & 262.9 $\pm$ 4.2 & 348.9 $\pm$
4.4\\
GRB020812   & 2257  & -15.30 & -12.13 & G & 1.3s & 30-400 & 20h38m48s
& -05d23m34s & 40.7 & -26.3 & 13.8$^{\prime}$ & 14.1 $\pm$ 0.6 & 42.0 $\pm$
1.0\\
GRB020813   & 2262  &   0.04 &  -3.81 & G & 1.3s & 30-400 & 19h46m38s
& -19d35m16s & 20.8 & -20.7 & 1$^{\prime}$(S) & $>$30.0 & $>$89.0\\
GRB020819   & 2275  &  17.70 & -22.45 & G & 160ms & 30-400 & 23h27m07s
& +06d21m50s & 88.5 & -50.8 & 7$^{\prime}$ & 11.5 $\pm$ 0.3 & 46.9 $\pm$ 2.0\\
GRB020903   & 2314  &   4.20 &  12.64 & XG & 5.12s & --- & 22h49m25s &
-20d53m59s & 38.9 & -61.5 & 16.7$^{\prime}$ & 4.8 $\pm$ 0.4 & 10.0 $\pm$ 0.7\\
GRB021004   & 2380  &   3.92 & -12.39 & G & 5.2s & 30-400 & 00h26m57s
& +18d55m44s & 114.9 & -43.6 & 2$^{\prime}$(S) & 26.6 $\pm$ 1.0 & 77.1 $\pm$
2.6\\
GRB021021   & 10623 &  15.24 &  11.92 & --- & --- & --- & 00h17m23s &
-01d37m00s & 103.8 & -63.2 & 20$^{\prime}$ & 22.1 $\pm$ 1.2 & 56.5 $\pm$ 1.2\\
GRB021104   & 2434  &  22.56 &  22.38 & G & 1.3s & 5-120 & 03h53m48s &
+37d57m12s & 158.1 & -12.2 & 26$^{\prime}$ & 10.2 $\pm$ 0.5 & 18.1 $\pm$ 0.2\\
GRB021112   & 2448  &  12.24 &  27.06 & G & 1.3s & 5-120 & 02h36m52s &
+48d50m56s & 140.2 & -10.5 & 20$^{\prime}$ & 6.8 $\pm$ 1.2 & 14.7 $\pm$ 1.1\\
GRB021211   & 2493  & -12.55 &  -0.01 & G & 160ms & 30-400 & 08h09m00s
& +06d44m20s & 215.7 & 20.3 & 2$^{\prime}$(S) & 3.1 $\pm$ 0.1 & 13.3 $\pm$
0.3\\
GRB030115   & 2533  &  13.01 &  -3.11 & G & 1.3s & 30-400 & 11h18m30s
& +15d02m17s & 237.4 & 65.2 & 2$^{\prime}$(S) & 9.2 $\pm$ 0.5 & 49.6 $\pm$
4.3\\
GRB030226   & 10893 & -13.00 & -16.27 & --- & --- & --- & 11h33m01s &
+25d53m56s & 212.5 & 72.4 & 2$^{\prime}$(S) & 66.4 $\pm$ 3.9 & 137.7 $\pm$
4.9\\
GRB030323   & 2640  &   4.05 &  35.06 & XG & 320ms & --- & 11h06m54s &
-21d51m00s & 273.0 & 34.9 & 18$^{\prime}$ & 13.9 $\pm$ 1.6 & 32.6 $\pm$ 2.7\\
GRB030324   & 2641  & -26.35 &   0.57 & G & 1.3s & 30-400 & 13h37m11s
& -00d19m22s & 326.6 & 60.4 & 7$^{\prime}$ & 8.9 $\pm$ 0.3 & 25.8 $\pm$ 0.8\\
GRB030328   & 2650  &   5.05 &   7.14 & G & 1.3s & 5-120 & 12h10m51s &
-09d21m05s & 286.4 & 52.2 & 1$^{\prime}$(S) & 106.9 $\pm$ 1.2 & 315.8 $\pm$
3.0\\
GRB030329   & 2652  &  26.68 & -29.00 & G & 1.3s & 5-120 & 10h44m49s &
+21d28m44s & 217.1 & 60.7 & 2$^{\prime}$(S) & 12.1 $\pm$ 0.2 & 33.1 $\pm$ 0.5\\
GRB030416   & 10897 &  -1.98 & -11.32 & --- & --- & --- & 11h06m51s &
-02d52m58s & 258.8 & 50.8 & 7$^{\prime}$ & 19.7 $\pm$ 1.7 & 61.5 $\pm$ 1.2\\
GRB030418   & 2686  &   7.45 &  -9.66 & XG & 13.280s & --- & 10h54m53s
& -06d59m22s & 259.1 & 45.7 & 9$^{\prime}$ & 38.7 $\pm$ 0.9 & 117.6 $\pm$ 0.7\\
\hline
\tablebreak

GRB030429   & 2695  &   8.88 &  11.83 & XG & 6.72s & --- & 12h13m06s &
-20d56m00s & 291.0 & 41.1 & 1$^{\prime}$(S) & 38.4 $\pm$ 1.5 & 77.4 $\pm$ 1.2\\
GRB030519   & 2716  & -41.00 &  16.18 & G & 160ms & 30-400 & 14h58m18s
& -32d56m57s & 331.5 & 22.8 & 30$^{\prime}$ & 6.1 $\pm$ 0.6 & 13.8 $\pm$ 0.7\\
GRB030528   & 2724  &  20.66 &   6.14 & G & 1.3s & 30-400 & 17h04m02s
& -22d38m59s & 0.0 & 11.3 & 2$^{\prime}$(S) & 20.8 $\pm$ 1.2 & 49.2 $\pm$ 1.2\\
GRB030723   & 2777  &   1.55 &  10.93 & XG & 6.72s & WXM & 21h49m30s &
-27d42m06s & 21.2 & -49.9 & 2$^{\prime}$(S) & 9.9 $\pm$ 0.3 & 20.2 $\pm$ 0.5\\
GRB030725   & 2779  &  18.41 &  33.10 & G & 160ms & 5-120 & 20h33m47s
& -50d45m49s & 348.2 & -36.6 & 14.4$^{\prime}$ & 68.3 $\pm$ 3.4 & 200.0 $\pm$
2.5\\
GRB030821   & 2814  &  12.13 &  32.47 & G & 1.3s & 30-400 & 21h42m33s
& -45d12m12s & 354.3 & -48.5 & 120$^{\prime}$x10$^{\prime}$ & 11.7 $\pm$ 1.5 &
22.9 $\pm$ 0.5\\
GRB030823   & 2818  &  11.67 & -32.65 & G & 5.2s & 5-120 & 21h30m47s &
+21d59m46s & 73.2 & -21.0 & 5.4$^{\prime}$ & 30.2 $\pm$ 1.4 & 66.4 $\pm$ 1.9\\
GRB030824   & 2821  & -29.79 & -31.43 & G & 1.3s & 5-120 & 00h05m02s &
+19d55m37s & 108.3 & -41.6 & 11.2$^{\prime}$ & 13.1 $\pm$ 1.8 & 36.4 $\pm$
0.4\\
GRB030913   & 2849  &  -2.05 &   4.62 & G & 1.3s & 30-400 & 20h58m02s
& -02d12m32s & 46.5 & -29.0 & 30$^{\prime}$ & 2.9 $\pm$ 0.3 & 6.7 $\pm$ 0.3\\
\enddata
\tablenotetext{a}{Triggered type; G: FREGATE triggered, XG: FREGATE triggered
by XDSP.}
\tablenotetext{b}{Trigger time-scale.}
\tablenotetext{c}{Trigger energy band in keV.}
\tablenotetext{d}{Location error radius (90\% confidence).  ``S'' indicates 
localization by the SXC.}
\tablenotetext{e}{Since the attitude control camera was not operational, 
the celestial coordinates is not available.}
\end{deluxetable}

\begin{table}
\caption{Spectral Parameters of 45 HETE-2 GRBs}
\vspace{0.5cm}
\label{hyo:hete_grb_spec_para}
\centerline{
{\scriptsize
\begin{tabular}{ccccccccc}\hline
GRB        & Class$^{a}$ & Model$^{b}$ & $\alpha$ & $\beta$ & $E_{\rm peak}$ [keV] & K$_{15}^{c}$ & $\chi^{2}_{\nu}$ & D.O.F.\\\hline\hline  
 GRB010213 & XRF & Band & $-1.00$(fixed) & $-2.96_{-0.54}^{+ 0.22}$ & $ 3.41_{-0.40}^{+ 0.35}$ & $44.63_{-5.19}^{+ 7.63}$ & 0.940 &  44\\
 GRB010225 & XRF &  PLE & $-1.31_{-0.26}^{+ 0.30}$ & --- & $31.57_{-9.17}^{+26.50}$ & $ 6.75_{-1.87}^{+ 2.93}$ & 0.925 &  39\\
GRB010326B$^{\dagger}$ & XRR &  PLE & $-1.08_{-0.22}^{+ 0.25}$ & --- & $51.77_{-11.25}^{+18.61}$ & $13.19_{-2.31}^{+ 3.07}$ & 0.856 & 111\\
 GRB010612 & GRB &  PLE & $-1.07_{-0.17}^{+ 0.19}$ & --- & $244.50_{-81.97}^{+285.07}$ & $ 2.93_{-0.36}^{+ 0.37}$ & 0.884 &  65\\
 GRB010613 & XRR & Band & $-0.95_{-0.26}^{+ 0.33}$ & $-2.01_{-0.15}^{+ 0.09}$ & $46.25_{-10.38}^{+17.78}$ & $15.24_{-2.43}^{+ 4.37}$ & 0.785 & 134\\
GRB010629B & XRR &  PLE & $-1.12_{-0.13}^{+ 0.14}$ & --- & $45.57_{-3.84}^{+ 4.61}$ & $20.05_{-1.56}^{+ 1.77}$ & 0.817 & 110\\
 GRB010921$^{\dagger}$ & XRR &  PLE & $-1.55_{-0.07}^{+ 0.08}$ & --- & $88.63_{-13.79}^{+21.76}$ & $41.79_{-1.63}^{+ 1.75}$ & 0.939 & 140\\
 GRB010928$^{\dagger}$ & GRB &  PLE & $-0.66_{-0.09}^{+ 0.10}$ & --- & $409.50_{-74.98}^{+118.46}$ & $ 6.30_{-0.55}^{+ 0.55}$ & 0.825 & 125\\
 GRB011019 & XRF &  PLE & $-1.43$ (fixed) & --- & $18.71_{-8.72}^{+18.33}$ & $ 2.46_{-0.63}^{+ 0.82}$ & 0.854 &  68\\
 GRB011103 & XRR &   PL & $-1.73_{-0.29}^{+ 0.24}$ & --- & --- & $ 2.72_{-0.88}^{+ 0.88}$ & 1.266 &  38\\
 GRB011130 & XRF &   PL & --- & $-2.65_{-0.33}^{+ 0.26}$ & $<$ 3.9$^{\rm d}$ & $ 0.69_{-0.26}^{+ 0.29}$ & 1.016 &  40\\
 GRB011212 & XRF &   PL & --- & $-2.07_{-0.22}^{+ 0.19}$ & --- & $ 0.72_{-0.18}^{+ 0.17}$ & 0.795 &  54\\
 GRB020124 & XRR &  PLE & $-0.79_{-0.14}^{+ 0.15}$ & --- & $86.93_{-12.45}^{+18.11}$ & $ 9.24_{-0.88}^{+ 0.98}$ & 0.710 &  95\\
 GRB020127 & XRR &  PLE & $-1.03_{-0.13}^{+ 0.14}$ & --- & $104.00_{-24.10}^{+47.00}$ & $ 4.50_{-0.51}^{+ 0.58}$ & 0.746 & 110\\
 GRB020317 & XRF &  PLE & $-0.61_{-0.52}^{+ 0.61}$ & --- & $28.41_{-7.41}^{+12.68}$ & $ 7.27_{-3.12}^{+ 7.73}$ & 0.923 &  53\\
 GRB020331 & GRB &  PLE & $-0.79_{-0.12}^{+ 0.13}$ & --- & $91.57_{-14.09}^{+20.99}$ & $ 4.03_{-0.41}^{+ 0.46}$ & 0.732 & 111\\
 GRB020531 & GRB &  PLE & $-0.83_{-0.13}^{+ 0.14}$ & --- & $230.60_{-58.11}^{+113.10}$ & $20.99_{-2.21}^{+ 2.31}$ & 0.831 & 141\\
 GRB020625 & XRF &  PLE & $-1.14$ (fixed) & --- & $ 8.52_{-2.91}^{+ 5.44}$ & $ 2.84_{-0.78}^{+ 1.05}$ & 0.781 &  55\\
 GRB020801$^{\dagger}$ & GRB & Band & $-0.32_{-0.34}^{+ 0.44}$ & $-2.01_{-0.25}^{+ 0.17}$ & $53.35_{-11.12}^{+14.37}$ & $ 5.66_{-1.02}^{+ 1.92}$ & 0.638 & 140\\
 GRB020812 & XRR &  PLE & $-1.09_{-0.25}^{+ 0.29}$ & --- & $87.62_{-29.57}^{+106.03}$ & $ 2.27_{-0.47}^{+ 0.61}$ & 0.664 &  68\\
 GRB020813$^{\dagger}$ & GRB & Band & $-0.94_{-0.03}^{+ 0.03}$ & $-1.57_{-0.04}^{+ 0.03}$ & $142.10_{-12.91}^{+14.05}$ & $20.74_{-0.47}^{+ 0.51}$ & 1.160 & 140\\
 GRB020819 & XRR & Band & $-0.90_{-0.14}^{+ 0.17}$ & $-1.99_{-0.48}^{+ 0.18}$ & $49.90_{-12.19}^{+17.88}$ & $10.71_{-1.65}^{+ 2.47}$ & 0.945 & 108\\
 GRB020903 & XRF &   PL & --- & $-2.62_{-0.55}^{+ 0.42}$ & $<$ 5.0$^{\rm d}$ ($2.6_{-0.8}^{+1.4})$ & $ 0.41_{-0.25}^{+ 0.34}$ & 0.845 &  26\\
 GRB021004 & XRR &  PLE & $-1.01_{-0.17}^{+ 0.19}$ & --- & $79.79_{-22.97}^{+53.35}$ & $ 2.77_{-0.48}^{+ 0.60}$ & 0.949 &  68\\
 GRB021021 & XRF &  PLE & $-1.33$ (fixed) & --- & $15.38_{-7.47}^{+14.24}$ & $ 1.24_{-0.37}^{+ 0.50}$ & 0.879 &  41\\
 GRB021104$^{\dagger}$ & XRF &  PLE & $-1.11_{-0.46}^{+ 0.54}$ & --- & $28.21_{-7.88}^{+17.41}$ & $ 7.59_{-2.55}^{+ 5.31}$ & 0.744 &  38\\
 GRB021112 & XRR &  PLE & $-0.94_{-0.32}^{+ 0.42}$ & --- & $57.15_{-20.70}^{+38.90}$ & $ 6.57_{-1.83}^{+ 3.47}$ & 1.126 &  61\\
 GRB021211 & XRR & Band & $-0.86_{-0.09}^{+ 0.10}$ & $-2.18_{-0.25}^{+ 0.14}$ & $45.56_{-6.23}^{+ 7.84}$ & $32.58_{-3.32}^{+ 4.16}$ & 1.149 & 140\\
 GRB030115 & XRR &  PLE & $-1.28_{-0.13}^{+ 0.14}$ & --- & $82.79_{-22.26}^{+52.82}$ & $ 3.50_{-0.46}^{+ 0.53}$ & 0.812 &  67\\
 GRB030226 & GRB &  PLE & $-0.89_{-0.15}^{+ 0.17}$ & --- & $97.12_{-17.06}^{+26.98}$ & $ 3.47_{-0.38}^{+ 0.42}$ & 0.894 & 139\\
 GRB030323 & XRR &   PL & $-1.62_{-0.25}^{+ 0.24}$ & --- & --- & $ 2.19_{-0.67}^{+ 0.64}$ & 0.835 &  33\\
 GRB030324 & XRR &  PLE & $-1.45_{-0.12}^{+ 0.14}$ & --- & $146.80_{-65.49}^{+627.57}$ & $ 4.94_{-0.60}^{+ 0.72}$ & 0.882 &  76\\
 GRB030328 & GRB & Band & $-1.14_{-0.03}^{+ 0.03}$ & $-2.09_{-0.40}^{+ 0.19}$ & $126.30_{-13.10}^{+13.89}$ & $ 6.64_{-0.18}^{+ 0.20}$ & 0.982 & 140\\
 GRB030329 & XRR & Band & $-1.26_{-0.02}^{+ 0.01}$ & $-2.28_{-0.06}^{+ 0.05}$ & $67.86_{-2.15}^{+ 2.31}$ & $146.20_{-1.70}^{+ 1.70}$ & 1.537 & 139\\
 GRB030416 & XRF &   PL & --- & $-2.31_{-0.15}^{+ 0.13}$ & $<$ 3.8$^{\rm d}$ (2.6$_{-1.8}^{+0.5}$) & $ 0.92_{-0.17}^{+ 0.17}$ & 0.870 &  54\\
 GRB030418 & XRR &  PLE & $-1.46_{-0.13}^{+ 0.14}$ & --- & $46.10_{-13.70}^{+31.96}$ & $ 2.43_{-0.37}^{+ 0.48}$ & 0.929 &  68\\
 GRB030429 & XRF &  PLE & $-1.12_{-0.22}^{+ 0.25}$ & --- & $35.04_{-7.90}^{+11.75}$ & $ 4.05_{-0.90}^{+ 1.32}$ & 0.720 &  68\\
 GRB030519$^{\dagger}$ & GRB & Band & $-0.75_{-0.06}^{+ 0.07}$ & $-1.72_{-0.07}^{+ 0.05}$ & $137.60_{-15.36}^{+17.80}$ & $73.21_{-1.90}^{+ 2.06}$ & 0.742 & 124\\
 GRB030528$^{\dagger}$ & XRF & Band & $-1.33_{-0.12}^{+ 0.15}$ & $-2.65_{-0.98}^{+ 0.29}$ & $31.84_{-4.97}^{+ 4.67}$ & $13.94_{-1.48}^{+ 2.44}$ & 0.809 & 109\\
 GRB030723 & XRF &   PL & --- & $-1.90_{-0.20}^{+ 0.16}$ & $<$ 8.9$^{\rm d}$ & $ 1.00_{-0.22}^{+ 0.21}$ & 0.952 &  142\\
 GRB030725 & XRR &  PLE & $-1.51_{-0.04}^{+ 0.04}$ & --- & $102.80_{-13.73}^{+19.05}$ & $15.71_{-0.50}^{+ 0.54}$ & 1.069 & 141\\
 GRB030821 & XRR &  PLE & $-0.88_{-0.12}^{+ 0.13}$ & --- & $84.26_{-10.88}^{+15.12}$ & $ 8.74_{-0.70}^{+ 0.77}$ & 0.971 &  98\\
 GRB030823 & XRF &  PLE & $-1.31_{-0.18}^{+ 0.20}$ & --- & $26.57_{-5.02}^{+ 7.45}$ & $ 8.26_{-1.59}^{+ 2.34}$ & 0.708 & 110\\
 GRB030824 & XRF &   PL & --- & $-2.14_{-0.14}^{+ 0.13}$ & $<$ 8.7$^{\rm d}$ (6.1$_{-4.2}^{+1.9}$) & $ 5.25_{-0.78}^{+ 0.76}$ & 0.813 &  53\\
 GRB030913 & GRB &  PLE & $-0.82_{-0.24}^{+ 0.28}$ & --- & $119.70_{-36.47}^{+113.25}$ & $ 3.53_{-0.62}^{+ 0.75}$ & 0.740 &  53\\\hline
\end{tabular}}
}
\vspace{0.2cm}
{\scriptsize
a GRB classification;  XRF: X-ray-flash, XRR: X-ray-rich GRB, GRB: GRB\vspace{-0.3cm}\\
b Spectral model; PL: Power-law; PLE: Power-law times exponential cutoff; Band: Band function
\vspace{-0.3cm}\\
c Normalization at 15 keV in units of 10$^{-2}$ photons cm$^{-2}$ s$^{-1}$ keV$^{-1}$\vspace{-0.3cm}\\ 
d 99.7\% upper limit and 90\% confidence interval (in parenthesis) derived by the 
{\it constrained} Band function\vspace{-0.3cm}\\
$\dagger$ The constant factor is multiplied to the spectral model.
}
\end{table}

\begin{table}
\caption{Photon Number and Photon Energy Fluences of 45 HETE-2 GRBs}
\vspace{0.5cm}
\label{hyo:hete_grb_photon_flux_fluence}
\centerline{
{\scriptsize
\begin{tabular}{c|c|rrr|rrr|c}\hline
GRB & Duration & \multicolumn{3}{c|}{Photon fluence$^{a}$} &
    \multicolumn{3}{|c|}{Energy fluence$^{b}$} & X/$\gamma$ ratio \\\hline 
    & [sec.]         & 2--30 keV & 30--400 keV & 2--400 keV & 2--30 
keV & 30--400 keV & 2--400 keV & \\\hline\hline
 GRB010213 & 34.41 & $111.10_{-4.80}^{+ 5.20}$ & $ 0.69_{-0.34}^{+ 0.69}$ & $111.80_{-4.80}^{+ 5.50}$ & $ 7.88_{-0.54}^{+ 0.25}$ & $ 0.69_{-0.32}^{+ 0.58}$ & $ 8.58_{-0.94}^{+ 1.01}$ & 11.38\\
 GRB010225 &  9.76 & $28.30_{-3.80}^{+ 3.71}$ & $ 2.73_{-0.88}^{+ 0.98}$ & $31.04_{-3.91}^{+ 3.90}$ & $ 3.48_{-0.36}^{+ 0.36}$ & $ 2.40_{-0.94}^{+ 1.72}$ & $ 5.89_{-1.06}^{+ 1.69}$ &  1.45\\
GRB010326B &  3.50 & $16.77_{-2.77}^{+ 2.83}$ & $ 3.25_{-0.56}^{+ 0.52}$ & $19.98_{-2.76}^{+ 2.91}$ & $ 2.40_{-0.27}^{+ 0.27}$ & $ 3.22_{-0.76}^{+ 0.92}$ & $ 5.62_{-0.81}^{+ 0.95}$ &  0.75\\
 GRB010612 & 47.19 & $56.63_{-11.80}^{+13.21}$ & $29.26_{-1.89}^{+ 2.36}$ & $85.89_{-12.27}^{+13.68}$ & $ 8.84_{-1.31}^{+ 1.35}$ & $50.23_{-6.21}^{+ 7.12}$ & $59.05_{-6.53}^{+ 6.37}$ &  0.18\\
 GRB010613 & 141.56 & $672.40_{-90.60}^{+111.80}$ & $168.50_{-5.70}^{+ 5.60}$ & $840.90_{-92.00}^{+111.80}$ & $101.50_{-6.54}^{+ 6.60}$ & $227.60_{-12.50}^{+12.60}$ & $329.40_{-13.80}^{+13.70}$ &  0.45\\
GRB010629B & 24.58 & $182.90_{-20.20}^{+21.40}$ & $29.99_{-1.72}^{+ 1.72}$ & $212.90_{-20.20}^{+21.30}$ & $25.41_{-1.66}^{+ 1.67}$ & $28.56_{-2.47}^{+ 2.69}$ & $53.97_{-3.13}^{+ 3.32}$ &  0.89\\
 GRB010921 & 23.85 & $610.10_{-45.80}^{+48.90}$ & $88.48_{-3.57}^{+ 3.58}$ & $698.80_{-45.50}^{+48.20}$ & $71.60_{-3.20}^{+ 3.42}$ & $112.60_{-8.40}^{+ 8.60}$ & $184.20_{-9.50}^{+ 9.70}$ &  0.64\\
 GRB010928 & 34.55 & $70.48_{-8.98}^{+ 9.68}$ & $104.70_{-3.10}^{+ 3.10}$ & $174.80_{-9.30}^{+10.00}$ & $13.71_{-1.35}^{+ 1.24}$ & $225.70_{-9.50}^{+ 9.10}$ & $239.30_{-4.80}^{+ 9.20}$ &  0.06\\
 GRB011019 & 24.57 & $27.52_{-5.65}^{+ 5.65}$ & $ 1.47_{-0.98}^{+ 1.23}$ & $28.99_{-5.65}^{+ 5.65}$ & $ 3.03_{-0.58}^{+ 0.58}$ & $ 1.10_{-0.74}^{+ 1.39}$ & $ 4.13_{-1.14}^{+ 1.54}$ &  2.77\\
 GRB011103 & 14.75 & $30.83_{-5.02}^{+ 5.01}$ & $ 4.13_{-2.21}^{+ 3.98}$ & $34.96_{-6.35}^{+ 6.78}$ & $ 3.31_{-0.71}^{+ 0.79}$ & $ 6.22_{-2.84}^{+ 8.72}$ & $ 9.53_{-4.30}^{+ 8.73}$ &  0.53\\
 GRB011130 & 50.00 & $86.00_{-13.00}^{+13.00}$ & $ 1.00_{-0.50}^{+ 1.00}$ & $87.00_{-13.00}^{+12.50}$ & $ 5.85_{-0.96}^{+ 0.98}$ & $ 0.98_{-0.62}^{+ 1.17}$ & $ 6.83_{-1.46}^{+ 1.90}$ &  5.96\\
 GRB011212 & 57.68 & $47.30_{-6.92}^{+ 6.92}$ & $ 2.31_{-0.58}^{+ 1.73}$ & $49.60_{-6.92}^{+ 7.50}$ & $ 4.24_{-0.64}^{+ 0.64}$ & $ 3.37_{-1.68}^{+ 2.53}$ & $ 7.61_{-2.16}^{+ 2.90}$ &  1.26\\
 GRB020124 & 40.63 & $115.00_{-11.80}^{+11.80}$ & $51.19_{-4.06}^{+ 3.66}$ & $165.80_{-12.20}^{+13.00}$ & $19.74_{-1.41}^{+ 1.40}$ & $61.33_{-7.63}^{+ 8.79}$ & $81.04_{-7.70}^{+ 8.86}$ &  0.32\\
 GRB020127 & 25.63 & $43.57_{-4.10}^{+ 4.10}$ & $15.63_{-1.53}^{+ 1.80}$ & $59.21_{-4.62}^{+ 4.61}$ & $ 6.73_{-0.50}^{+ 0.51}$ & $20.49_{-3.65}^{+ 4.48}$ & $27.22_{-3.63}^{+ 4.43}$ &  0.33\\
 GRB020317 & 10.00 & $13.80_{-3.30}^{+ 3.50}$ & $ 1.70_{-0.70}^{+ 0.80}$ & $15.50_{-3.40}^{+ 3.70}$ & $ 2.16_{-0.38}^{+ 0.37}$ & $ 1.29_{-0.64}^{+ 0.88}$ & $ 3.45_{-0.79}^{+ 0.94}$ &  1.68\\
 GRB020331 & 75.00 & $93.00_{-6.75}^{+ 7.50}$ & $43.50_{-3.75}^{+ 3.75}$ & $136.50_{-8.30}^{+ 8.30}$ & $16.07_{-1.03}^{+ 1.04}$ & $53.32_{-7.39}^{+ 8.52}$ & $69.40_{-7.45}^{+ 8.45}$ &  0.30\\
 GRB020531 &  1.04 & $ 7.54_{-1.09}^{+ 1.12}$ & $ 6.18_{-0.48}^{+ 0.48}$ & $13.72_{-1.21}^{+ 1.22}$ & $ 1.33_{-0.14}^{+ 0.14}$ & $10.96_{-1.34}^{+ 1.35}$ & $12.30_{-1.35}^{+ 1.35}$ &  0.12\\
 GRB020625 & 41.94 & $24.74_{-3.77}^{+ 4.20}$ & $ 0.19_{-0.18}^{+ 0.51}$ & $25.16_{-3.77}^{+ 3.78}$ & $ 2.37_{-0.50}^{+ 0.55}$ & $ 0.12_{-0.11}^{+ 0.35}$ & $ 2.49_{-0.58}^{+ 0.83}$ & 20.49\\
 GRB020801 & 117.97 & $130.90_{-23.50}^{+27.20}$ & $69.60_{-4.72}^{+ 4.72}$ & $200.50_{-23.50}^{+27.20}$ & $25.66_{-2.72}^{+ 2.83}$ & $95.23_{-10.54}^{+10.67}$ & $121.00_{-10.90}^{+11.00}$ &  0.27\\
 GRB020812 & 60.16 & $52.94_{-9.62}^{+10.23}$ & $15.64_{-3.01}^{+ 3.61}$ & $68.58_{-9.62}^{+10.83}$ & $ 7.91_{-1.09}^{+ 1.09}$ & $19.15_{-5.86}^{+ 8.18}$ & $27.06_{-5.98}^{+ 8.11}$ &  0.41\\
 GRB020813 & 113.00 & $845.20_{-21.40}^{+22.60}$ & $480.20_{-4.50}^{+ 4.60}$ & $1325.00_{-22.00}^{+24.00}$ & $138.50_{-2.30}^{+ 2.70}$ & $839.60_{-12.50}^{+12.30}$ & $978.70_{-12.80}^{+12.70}$ &  0.16\\
 GRB020819 & 50.16 & $163.00_{-9.00}^{+ 8.50}$ & $45.65_{-3.01}^{+ 3.51}$ & $208.70_{-9.60}^{+ 9.50}$ & $25.20_{-1.11}^{+ 1.10}$ & $62.53_{-9.27}^{+ 8.33}$ & $87.80_{-9.47}^{+ 8.39}$ &  0.40\\
 GRB020903 & 13.00 & $12.61_{-2.60}^{+ 2.73}$ & $ 0.16_{-0.13}^{+ 0.34}$ & $12.74_{-2.73}^{+ 2.73}$ & $ 0.83_{-0.24}^{+ 0.28}$ & $ 0.16_{-0.14}^{+ 0.44}$ & $ 0.95_{-0.33}^{+ 0.62}$ &  7.31\\
 GRB021004 & 49.70 & $49.70_{-4.47}^{+ 4.97}$ & $15.41_{-2.98}^{+ 2.48}$ & $65.11_{-5.47}^{+ 5.46}$ & $ 7.65_{-0.69}^{+ 0.69}$ & $17.79_{-5.00}^{+ 7.01}$ & $25.45_{-5.04}^{+ 6.85}$ &  0.43\\
 GRB021021 & 49.15 & $23.10_{-5.41}^{+ 5.41}$ & $ 0.88_{-0.69}^{+ 1.13}$ & $24.08_{-5.89}^{+ 5.41}$ & $ 2.51_{-0.64}^{+ 0.63}$ & $ 0.62_{-0.49}^{+ 1.07}$ & $ 3.13_{-1.06}^{+ 1.39}$ &  4.03\\
 GRB021104 & 31.41 & $78.84_{-21.05}^{+29.86}$ & $ 7.54_{-2.51}^{+ 3.14}$ & $86.38_{-21.36}^{+30.12}$ & $10.32_{-1.80}^{+ 2.06}$ & $ 6.13_{-2.67}^{+ 4.40}$ & $16.38_{-3.34}^{+ 5.00}$ &  1.69\\
 GRB021112 &  4.00 & $ 8.48_{-2.00}^{+ 2.08}$ & $ 2.12_{-0.64}^{+ 0.64}$ & $10.60_{-2.12}^{+ 2.20}$ & $ 1.31_{-0.25}^{+ 0.25}$ & $ 2.14_{-0.90}^{+ 1.08}$ & $ 3.45_{-0.93}^{+ 1.09}$ &  0.61\\
 GRB021211 &  8.00 & $74.56_{-2.40}^{+ 2.40}$ & $18.88_{-0.88}^{+ 0.88}$ & $93.44_{-2.56}^{+ 2.64}$ & $11.60_{-0.29}^{+ 0.29}$ & $23.71_{-2.01}^{+ 2.03}$ & $35.34_{-2.06}^{+ 2.07}$ &  0.49\\
 GRB030115 & 36.00 & $58.68_{-5.40}^{+ 5.40}$ & $12.60_{-1.80}^{+ 1.44}$ & $71.28_{-5.76}^{+ 5.76}$ & $ 7.89_{-0.61}^{+ 0.61}$ & $15.17_{-3.20}^{+ 4.02}$ & $23.05_{-3.23}^{+ 3.98}$ &  0.52\\
 GRB030226 & 68.81 & $79.82_{-10.32}^{+10.32}$ & $33.72_{-2.76}^{+ 3.44}$ & $114.20_{-11.00}^{+10.30}$ & $13.20_{-1.18}^{+ 1.18}$ & $42.92_{-6.02}^{+ 6.85}$ & $56.12_{-6.14}^{+ 6.93}$ &  0.31\\
 GRB030323 & 19.61 & $29.22_{-12.55}^{+16.08}$ & $ 5.49_{-1.76}^{+ 1.57}$ & $34.71_{-12.55}^{+15.49}$ & $ 3.40_{-1.21}^{+ 1.29}$ & $ 8.91_{-3.48}^{+ 3.84}$ & $12.30_{-3.43}^{+ 3.68}$ &  0.38\\
 GRB030324 & 15.73 & $43.73_{-4.25}^{+ 4.56}$ & $ 8.97_{-1.26}^{+ 1.25}$ & $52.70_{-4.57}^{+ 4.71}$ & $ 5.49_{-0.44}^{+ 0.44}$ & $12.75_{-3.01}^{+ 3.35}$ & $18.23_{-3.01}^{+ 3.34}$ &  0.43\\
 GRB030328 & 199.23 & $555.90_{-10.00}^{+11.90}$ & $193.30_{-4.00}^{+ 5.90}$ & $751.10_{-12.00}^{+12.00}$ & $81.86_{-1.29}^{+ 1.31}$ & $287.40_{-14.10}^{+13.90}$ & $369.50_{-14.20}^{+14.00}$ &  0.28\\
 GRB030329 & 62.94 & $4121.00_{-42.00}^{+41.00}$ & $843.40_{-5.70}^{+ 5.70}$ & $4963.00_{-40.00}^{+43.00}$ & $553.10_{-3.20}^{+ 3.10}$ & $1076.00_{-14.00}^{+13.00}$ & $1630.00_{-13.00}^{+14.00}$ &  0.51\\
 GRB030416 & 78.64 & $114.00_{-10.20}^{+ 9.50}$ & $ 3.15_{-0.79}^{+ 1.57}$ & $117.20_{-10.20}^{+10.20}$ & $ 8.98_{-0.87}^{+ 0.87}$ & $ 3.72_{-1.38}^{+ 1.85}$ & $12.70_{-2.09}^{+ 2.47}$ &  2.42\\
 GRB030418 & 110.10 & $143.10_{-8.80}^{+ 8.80}$ & $16.51_{-3.30}^{+ 4.41}$ & $160.70_{-9.90}^{+ 8.90}$ & $17.11_{-1.10}^{+ 1.09}$ & $17.34_{-5.22}^{+ 7.27}$ & $34.45_{-5.48}^{+ 7.23}$ &  0.99\\
 GRB030429 & 24.58 & $35.15_{-4.43}^{+ 4.67}$ & $ 4.42_{-1.23}^{+ 0.98}$ & $39.57_{-4.67}^{+ 4.67}$ & $ 4.74_{-0.49}^{+ 0.49}$ & $ 3.80_{-1.17}^{+ 1.40}$ & $ 8.54_{-1.32}^{+ 1.48}$ &  1.25\\
 GRB030519 & 20.97 & $485.70_{-23.90}^{+24.50}$ & $357.50_{-4.40}^{+ 2.30}$ & $843.20_{-23.50}^{+24.30}$ & $87.05_{-2.38}^{+ 2.43}$ & $609.30_{-9.70}^{+ 9.70}$ & $696.70_{-9.90}^{+ 9.90}$ &  0.14\\
 GRB030528 & 83.56 & $512.20_{-39.30}^{+40.10}$ & $53.48_{-3.34}^{+ 3.34}$ & $565.70_{-38.40}^{+40.90}$ & $62.54_{-2.79}^{+ 2.80}$ & $56.34_{-7.32}^{+ 7.13}$ & $119.00_{-7.80}^{+ 7.60}$ &  1.11\\
 GRB030723 & 31.25 & $28.70_{-4.15}^{+ 4.18}$ & $ 0.58_{-0.49}^{+ 3.30}$ & $29.27_{-4.45}^{+ 7.49}$ & $ 2.84_{-0.50}^{+ 0.49}$ & $ 0.38_{-0.33}^{+ 5.56}$ & $ 3.23_{-0.76}^{+ 5.82}$ &  7.47\\
 GRB030725 & 83.88 & $785.10_{-27.70}^{+27.70}$ & $126.70_{-4.20}^{+ 4.20}$ & $911.80_{-27.70}^{+28.50}$ & $94.12_{-2.25}^{+ 2.27}$ & $166.70_{-10.10}^{+10.30}$ & $260.80_{-10.40}^{+10.60}$ &  0.56\\
 GRB030821 & 21.21 & $60.66_{-5.73}^{+ 6.15}$ & $23.12_{-1.49}^{+ 1.48}$ & $83.78_{-5.94}^{+ 6.36}$ & $ 9.96_{-0.64}^{+ 0.63}$ & $27.47_{-2.99}^{+ 3.35}$ & $37.43_{-3.07}^{+ 3.41}$ &  0.36\\
 GRB030823 & 55.56 & $191.10_{-17.20}^{+18.40}$ & $15.56_{-3.34}^{+ 3.33}$ & $206.70_{-18.40}^{+18.30}$ & $23.05_{-1.55}^{+ 1.56}$ & $12.74_{-3.53}^{+ 4.43}$ & $35.80_{-3.97}^{+ 4.63}$ &  1.81\\
 GRB030824 & 15.73 & $103.30_{-15.05}^{+15.30}$ & $ 4.72_{-1.42}^{+ 1.42}$ & $107.90_{-14.78}^{+15.10}$ & $ 8.90_{-1.07}^{+ 1.07}$ & $ 5.83_{-1.89}^{+ 2.38}$ & $14.73_{-2.42}^{+ 2.72}$ &  1.53\\
 GRB030913 &  9.12 & $10.49_{-1.83}^{+ 1.82}$ & $ 5.84_{-0.82}^{+ 0.82}$ & $16.32_{-2.09}^{+ 2.01}$ & $ 1.81_{-0.23}^{+ 0.23}$ & $ 8.04_{-1.93}^{+ 2.69}$ & $ 9.86_{-1.73}^{+ 2.66}$ &  0.23\\\hline
\end{tabular}}
}
\vspace{0.2cm}
{\scriptsize
a Photon number fluences are in units of photon cm$^{-2}$ \vspace{-0.3cm}\\
b Photon energy fluences are in units of 10$^{-7}$ erg cm$^{-2}$ \vspace{-0.3cm}\\
}
\end{table}

\begin{table}
\caption{Peak (1 s) Photon Number Flux of 45 HETE-2 GRBs}
\vspace{0.5cm}
\label{hyo:hete_grb_spec_para}
\centerline{
{\scriptsize
\begin{tabular}{ccccccc}\hline
GRB        & Class$^{a}$ & Model$^{b}$ & F$^{N}_{\rm peak (2-30 \,
 keV)}$$^{c}$ & F$^{N}_{\rm peak (30-400 \, keV)}$ & F$^{N}_{\rm peak (2-400 \, keV)}$ & F$^{N}_{\rm peak (50-300 \, keV)}$\\\hline\hline
 GRB010213 & XRF & Band  & 6.33  $\pm$  0.77 &  (2.97 $\pm$ 0.55) $\times$ 10$^{-3}$ &  6.33 $\pm$  0.81 & (1.08 $\pm$ 0.15) $\times$ 10$^{-2}$\\
 GRB010225 & XRF &  PLE  & 5.11  $\pm$  2.36 &  0.33 $\pm$  0.17                     &  5.45 $\pm$  2.39 & (9.56 $\pm$ 9.37) $\times$ 10$^{-2}$\\
GRB010326B & XRR &  PLE  & 10.52 $\pm$  3.29 &  1.51 $\pm$  0.35                     & 12.03 $\pm$  3.33 &  0.73 $\pm$  0.24\\
 GRB010612 & GRB &  PLE  & 4.32  $\pm$  1.16 &  4.35 $\pm$  0.48                     &  8.67 $\pm$  1.38 &  3.07 $\pm$  0.35\\
 GRB010613 & XRR & Band  & 24.66 $\pm$ 11.60 &  2.68 $\pm$  0.87                     & 27.34 $\pm$ 11.19 &  1.23 $\pm$  0.40\\
GRB010629B & XRR &  PLE  & 39.08 $\pm$  7.30 &  4.17 $\pm$  0.42                     & 43.25 $\pm$  7.40 &  1.86 $\pm$  0.26\\
 GRB010921 & XRR &  PLE  & 34.20 $\pm$  4.05 &  5.74 $\pm$  0.46                     & 39.93 $\pm$  4.21 &  3.19 $\pm$  0.30\\
 GRB010928 & GRB &  PLE  & 3.83  $\pm$  0.76 &  6.91 $\pm$  0.45                     & 10.74 $\pm$  0.94 &  5.02 $\pm$  0.46\\
 GRB011019 & XRF &   PL  & 3.62  $\pm$  1.41 &  0.15 $\pm$  0.13                     &  3.76 $\pm$  1.44 & (7.41 $\pm$ 7.33) $\times$ 10$^{-2}$\\
 GRB011103 & XRR &   PL  & 4.42  $\pm$  1.12 &  0.14 $\pm$  0.08                     &  4.55 $\pm$  1.14 & (6.53 $\pm$ 4.27) $\times$ 10$^{-2}$\\
 GRB011130 & XRF &   PL  & 5.27  $\pm$  1.27 &  (8.20 $\pm$ 6.28)  $\times$ 10$^{-2}$ &  5.35 $\pm$  1.28 & (3.57 $\pm$ 3.21) $\times$ 10$^{-2}$\\
 GRB011212 & XRF &   PL  & 1.13  $\pm$  0.97 &  $<$ 7.66 $\times$ 10$^{-2}$ &  1.14 $\pm$  0.96 & $<$4.43 $\times$ 10$^{-2}$\\
 GRB020124 & XRR &  PLE  & 6.90  $\pm$  1.63 &  2.49 $\pm$  0.40 &  9.38 $\pm$  1.77 &  1.43 $\pm$  0.28\\
 GRB020127 & XRR &  PLE  & 5.95  $\pm$  1.17 &  2.17 $\pm$  0.42 &  8.12 $\pm$  1.50 &  1.27 $\pm$  0.27\\
 GRB020317 & XRF &  PLE  & 4.63  $\pm$  1.04 &  0.64 $\pm$  0.25 &  5.26 $\pm$  1.13 &  0.20 $\pm$  0.14\\
 GRB020331 & GRB &  PLE  & 1.93  $\pm$  0.37 &  1.72 $\pm$  0.23 &  3.65 $\pm$  0.51 &  1.19 $\pm$  0.17\\
 GRB020531 & GRB &  PLE  & 17.41 $\pm$  4.46 &  5.56 $\pm$  0.74 & 22.97 $\pm$  4.69 &  3.58 $\pm$  0.51\\
 GRB020625 & XRF &   PL  & 2.86  $\pm$  0.97 &  0.31 $\pm$  0.17 &  3.17 $\pm$  1.07 &  0.18 $\pm$  0.10\\
 GRB020801 & GRB & Band  & 6.36  $\pm$  1.13 &  1.38 $\pm$  0.25 &  7.73 $\pm$  2.11 &  0.79 $\pm$  0.18\\
 GRB020812 & XRR &  PLE  & 2.48  $\pm$  0.84 &  0.84 $\pm$  0.26 &  3.32 $\pm$  1.00 &  0.47 $\pm$  0.17\\
 GRB020813 & GRB & Band  & 19.53 $\pm$  1.29 & 12.79 $\pm$  0.83 & 32.31 $\pm$  2.07 &  8.31 $\pm$  0.55\\
 GRB020819 & XRR & Band  & 12.09 $\pm$  1.05 &  5.60 $\pm$  0.44 & 17.68 $\pm$  1.34 &  3.42 $\pm$  0.29\\
 GRB020903 & XRF &   PL  & 2.75  $\pm$  0.66 &  3.23$_{-2.40}^{+6.73}$ $\times$ 10$^{-2}$  &  2.78 $\pm$  0.67 & 1.37$_{-1.07}^{+3.68}$ $\times$ 10$^{-2}$\\
 GRB021004 & XRR &  PLE  & 1.80  $\pm$  0.38 &  0.89 $\pm$  0.20 &  2.69 $\pm$  0.50 &  0.43 $\pm$  0.15\\
 GRB021021 & XRF &  PL   & 2.14  $\pm$  1.06 &  0.31 $\pm$  0.24 &  2.45 $\pm$  1.17 &  0.19 $\pm$  0.16\\
 GRB021104 & XRF &  PLE  & 4.23  $\pm$  1.79 &  0.67 $\pm$  0.22 &  4.89 $\pm$  1.83 &  0.25 $\pm$  0.13\\
 GRB021112 & XRR &  PLE  & 3.45  $\pm$  1.15 &  1.03 $\pm$  0.37 &  4.47 $\pm$  1.29 &  0.55 $\pm$  0.28\\
 GRB021211 & XRR & Band  & 21.60 $\pm$  1.33 &  8.36 $\pm$  0.56 & 29.97 $\pm$  1.74 &  4.10 $\pm$  0.34\\
 GRB030115 & XRR &  PLE  & 6.97  $\pm$  1.32 &  1.16 $\pm$  0.17 &  8.13 $\pm$  1.38 &  1.16 $\pm$  0.17\\
 GRB030226 & GRB &  PLE  & 1.71  $\pm$  0.51 &  0.99 $\pm$  0.17 &  2.69 $\pm$  0.57 &  0.63 $\pm$  0.14\\
 GRB030323 & XRR &   PL  & 3.37  $\pm$  2.10 &  0.49 $\pm$  0.22 &  3.86 $\pm$  2.11 &  0.29 $\pm$  0.15\\
 GRB030324 & XRR &  PLE  & 6.63  $\pm$  1.04 &  1.63 $\pm$  0.30 &  8.27 $\pm$  1.20 &  0.96 $\pm$  0.23\\
 GRB030328 & GRB & Band  & 6.72  $\pm$  0.51 &  4.92 $\pm$  0.33 & 11.64 $\pm$  0.85 &  3.32 $\pm$  0.24\\
 GRB030329 & XRR & Band  & 378.59 $\pm$ 21.20 & 72.20 $\pm$  3.77 & 450.88 $\pm$ 24.68 & 38.06 $\pm$  2.04\\
 GRB030416 & XRF &   PL  & 4.50  $\pm$  0.91 &  0.26 $\pm$  0.10 & 4.77 $\pm$  0.94 &  (1.39 $\pm$  0.62) $\times$ 10$^{-2}$\\
 GRB030418 & XRR &  PLE  & 3.69  $\pm$  0.85 &  0.30 $\pm$  0.15 &  3.99 $\pm$  0.91 &  0.13 $\pm$  0.10\\
 GRB030429 & XRF &  PLE  & 3.08  $\pm$  0.72 &  0.71 $\pm$  0.19 &  3.79 $\pm$  0.79 &  0.29 $\pm$  0.11\\
 GRB030519 & GRB & Band  & 7.52  $\pm$  3.37 & 11.89 $\pm$  4.81 & 19.41 $\pm$  7.96 &  8.36 $\pm$  3.38\\
 GRB030528 & XRF & Band  & 17.28 $\pm$  1.52 &  0.61 $\pm$  0.12 & 17.89 $\pm$  1.57 &  (1.50 $\pm$  0.55) $\times$ 10$^{-1}$\\
 GRB030723 & XRF &  PLE  & 1.98  $\pm$  0.38 &  0.12$_{-0.09}^{+0.14}$ & 2.10 $\pm$  0.41 &  3.06$_{-2.57}^{+9.37}$ $\times$ 10$^{-2}$\\
 GRB030725 & XRR &  PLE  & 24.83 $\pm$  1.79 &  9.12 $\pm$  0.55 & 33.96 $\pm$  2.15 &  5.69 $\pm$  0.37\\
 GRB030821 & XRR &  PLE  & 3.84  $\pm$  0.72 &  1.93 $\pm$  0.27 &  5.77 $\pm$  0.86 &  1.19 $\pm$  0.19\\
 GRB030823 & XRF &  PLE  & 7.03  $\pm$  1.62 &  0.57 $\pm$  0.26 &  7.60 $\pm$  1.70 &  0.17 $\pm$  0.14\\
 GRB030824 & XRF &   PL  & 12.37 $\pm$  3.77 &  0.28 $\pm$  0.14 & 12.65 $\pm$  3.82 &  (1.29 $\pm$  0.74) $\times$ 10$^{-1}$\\
 GRB030913 & GRB &  PLE  & 2.20 $\pm$  0.48 &  1.36 $\pm$  0.25 &  3.55 $\pm$  0.63 &  0.89 $\pm$  0.18\\\hline
\end{tabular}}
}
\vspace{0.2cm}
{\scriptsize
a GRB classification;  XRF: X-ray-flash, XRR: X-ray-rich GRB, GRB: GRB \vspace{-0.3cm}\\
b Spectral model; PL: power-law, PLE: power-law times exponential cutoff, Band: Band function\vspace{-0.3cm}\\
c Photon number peak fluxes are in units of photons cm$^{-2}$s$^{-1}$\vspace{-0.3cm}\\
}
\end{table}

\acknowledgments



\clearpage

\clearpage

\clearpage

\end{document}